\documentclass[onecolumn,aps, tightpreprint,showpacs]{revtex4}
\usepackage{graphicx}
\usepackage{epsfig}
 \begin{document}
\setlength{\unitlength}{1mm}
\bibliographystyle{unsrt} 
\title{ Entropic Elasticity of Double-Strand DNA  Subject to  Simple Spatial Constraints  }
 \author{C. Bouchiat}
\affiliation{Laboratoire de Physique Th\'eorique de l'Ecole Normale Sup\'erieure \\ 
24, rue Lhomond, F-75231 Paris Cedex 05, France.}
\date {\today}
\begin{abstract}
The aim  of the present paper  is  the  study of the entropic  elasticity  of  the dsDNA molecule,
having a cristallographic length  $ L$ of the order of 10 to 30 persistence lengths $ A$,  when it is  subject 
to  spatial  obstructions. 
We have  not tried to obtain the single molecule partition function by solving 
 a Sch\"odringer-like equation. We  prefer   to stay within a 
discretized version of the WLC model with an  added one-monomer potential, 
simulating the spatial constraints. We  derived directly from the 
 discretized Boltzmann formula the transfer matrix connecting the  
partition functions  relative  to adjacent ``effective  monomers". 
We have plugged   adequate Dirac $\delta$-functions in the functional integral 
to ensure that  the monomer coordinate  and  the tangent vector  are
independent variables. The partition function is, then, given by an  iterative 
process which is both numerically  efficient and physically transparent.
 As a  test  of our discretized approach, we have studied 
two   configurations  involving  a dsDNA molecule  confined between 
a pair of   parallel plates. One  molecule  end  is anchored to  one   plate by a biochemical bond. A stretching  force $ F$,
normal to the plates, is pulling away the other end. In the first case, 
the cristallographic length $L $ is smaller than the two-plate distance $L_0$.
 The molecule feels,  then, only the anchoring 
barrier effect. The predicted  elongation-versus-force  curve, is pushed upward with respect to the  WLC  model result.
This effect is the most  spectacular in the low force  regime. For large forces say, $F_{high}  = 5\, k_B \, T/ A$, the elongation  
versus $L$ is  very well fitted by a straight line with  a slope given by the  standard WLC model and  a constant term $\simeq  1.2 A$.  
In  the second  case,  $L$   takes values up to   $ L_{max}= 1.5 \,L_0$. 
With a stretching  force still equal to  $F_{high}$, the  standard WLC model   predicts that the   molecule cannot fit  within the plates 
 when  $L >L_*=1.29 L_0$. We have studied the evolution of   the elongation derivative with
respect  to $L$, together with   the  mean square free-end fluctuations along the force. They both exibit   a sharp decrease when 
$ L \geq L_0$. We present a semi-qantitative argument suggesting
that the terminal segment  involving   $ 20 \,\% $ of the internal monomers flattens  
against the  repulsive barrier  when $ L  \rightarrow  L_{max}$. In conclusion, we suggest  extensions of the present work,
 relevant to the analysis of micromanipulation experiments. Finally, we have gathered into the Appendix 
formal developments,   leading to a precise relation between the transfer matrix and the Hamiltonian methods  for the study 
of spatially constrained dsDNA.

 \end{abstract}
\maketitle
%%%%%%%%%%%%%%%%%%%
  \newcommand \be {\begin{equation}}
\newcommand \ee {\end{equation}}
 \newcommand \bea {\begin{eqnarray}}
\newcommand \eea {\end{eqnarray}}
\newcommand \nn \nonumber
\def \(({\left(}
\def \)){\right)}
 \def \vr{{\mathbf{r}}}
\def \vv{{\mathbf{v}}}
 \def \vk{{\mathbf{k}}}
\def \vq{{\mathbf{q}}}
\def \vf{{\mathbf{f}}}
\def \vt{{\mathbf{t}}}
\def \vu{{\mathbf{u}}}
\def \vp{{\mathbf{p}}}
\def \vR{{\mathbf{R}}}  
 \def \va{{\mathbf{a}}}
 \def \vb{{\mathbf{b}}}
\def\bra{\langle}
\def\ket{\rangle}
%%%%%%%%%%%%%%%%%% 
\section*{Introduction  }
       In the last ten years Single Molecule Biophysics has become a very active field of research. Among the recently explored areas,
one finds the observation,  at the one-molecule  level, of the biochemical interactions  of the double strand  DNA  (dsDNA) with the various 
proteins involved in the duplication process (for two recent reviews see, for instance, the references \cite{busta,allem}). The
protein-DNA interaction  is detected  by the observation, in real time, of the variations 
 of the dsDNA elongation under the action of a fixed stretching force. It was clearly of interest to  have  a  good physical
understanding of the dsDNA elasticity measurements \cite{smith,perk,strick}. In absence of  DNA supercoiling,  this is  provided  by the so
called ``Worm Like Chain"  (WLC) model 
\cite{fixman,marsig,bouchbiophys} which gives a good description of the dsDNA  
entropic bending elasticity   within  a wide range of force, 
from few hundredths  to  few tens of picoNewton.  

In practice, the persistence length of a double helix  is about five  times  the  typical  length  resolution. It is, then, legitimate to use 
   a rectifiable curve  to represent the  coarse-grained  dsDNA  chain. 
The basic hypothesis of the  WLC  model   is  to  assume   that the
elastic-energy linear density is inversely proportional to the square of the chain-curvature radius.   In the usual  formulation of
the  WLC  model, the relevant Statistical  Mechanics variable is the  tangent vector $ \vt(s)=  d\,\vr /  d\,s $, where $\vr(s) $ is
the effective monomer coordinate and  $ds$ 
 the rectifiable chain length element. The elastic molecular-chain energy, expressed in thermal  unit 
   $  k_B\,T$,   is given by the following line integral:
\be
  E_{WLC}= \int_0^L d\,s  \(( \;\frac{A}{2}\, (\frac{ d\, \vt(s)} { ds} )^2- \vf  \cdot  \vt(s)  \)),
\label{energysimflex}
\ee  
where $L$ is the crystallographic length of the polymer.  $A $ is the  persistence length and $(d\, \vt(s) / ds )^2$
 is the  inverse  square of  the  curvature radius  of the coarse-grained  chain.  $  {\bf{ F}}=k_B\,T \,\vf   $   stands for  the
stretching force,  applied to the extremity of  the chain. It follows from its very definition  that $\vt$ is a unit vector. The
partition function $ Z$  is given by the functional Boltzmann integral:
\be
Z = \int {\cal{D}}\ [ \,\theta\ ]  {\cal{D}}\ [\,\phi\ ]  \exp\((-E_{WLC}\)) \, ,
\ee
where  the integration has to be performed over all the paths joining two points on  the unit sphere.
Exploiting the analogy with Feynman formulation of Quantum Mechanics (QM),  
the result  of the functional integration is given in terms of the  Hamiltonian operator \cite{saito}:
\be
{\widehat{H}}_{WLC}=  -\frac{1}{2\, A} {\nabla}_{\vt}^2 - \vf \cdot \vt, 
\ee
where ${\nabla}_{\vt}^2$ is the Laplacian on the unit sphere.
In the limit where the $L\gg A$, the elongation 
of the molecule under the action of a stretching force $ F =k_B \,  T \, f$  directed along the $z-axis$
 is given by:
\be
\bra\, z(L) \, \ket = -L\, \frac{\partial\, E_0(f)}{ \partial \, f}, 
\ee
  where  $E_0(f)$, the ground state eigen-energy  of  the Hamiltonian ${\widehat{H}}_{WLC}$, can be 
obtained with a very good precision by solving numerically an ordinary differential equation with 
appropriate boundary conditions  \cite{bouchbiophys}.
%%%%%%%%%%%%%%%%%%%%%%%%%%%%%%%%%%%%
\subsection{ The problem of the Spatial  Constraints in the WLC Model.}
%%%%%%%%%%%%%%%%%%%%%%%%%%%%%%%%%%%%%
 However, this approach becomes
ackward if one wishes to impose physical constraints involving the monomer space coordinate. For instance, one may wish
to confine the molecule within a certain region of space. This constraint can be formulated in terms of a potential, $V(\vr(s))$,
{\it acting on each monomer of coordinate $\vr(s)$}. 
If the   unitary tangent vector   $\vt(s)$ is the sole dynamical variable, one has to write
$\vr(s) =\int_0^s\vt(s') ds'$ and the potential energy to be added to $E_{WLC}$ has the  non-local form:
\be
\Delta E^{pot}_{WLC} = \int_0^L ds\;  V \left( \int_0^s \vt(s') ds' \right)\, .
\ee
 This situation becomes worse if one studies self-avoiding effects described by a monomer-monomer repulsive  potential, $V\left(
 \vr(s_1), \vr(s_2)\right)$:
  \be
\Delta E^{SA}_{WLC}=  \hspace{-2mm} \int_0^L \hspace{-1mm}ds_1  \hspace{-1mm} \int_0^L  \hspace{-1mm}ds_2 \;
 V\left(
 \hspace{-1mm}
 \int_0^{s_1}  \hspace{-1mm} \vt(s'_1) ds'_1 \, , 
 \hspace{-1mm}\int_0^{s_2}  \hspace{-1mm}\vt(s'_2) ds'_2 \right)\,.
 \ee

The  obvious thing   to avoid  this problem for the monomer 
potential energy  is  to use  the  monomer coordinate $ \vr(s)$ as a dynamical variable.
It is convenient  to  consider  a  whole family of polymer 
models  defined by the  elastic-energy linear density:
\be
 {\cal{E}}(s) ={\cal{E}}_0(\dot{\vr}^2 )+ \frac{1}{2} \,A\,{ \ddot {\vr} }^2  - \vf \cdot \dot {\vr}  + V(\vr),
\label{elasdens}
\ee
where the WLC model can be recovered by an appropriate choice of   ${\cal{E}}_0(\dot{\vr}^2 )$. 
But there is, clearly, a price to be  paid :
  second-order derivatives appear now  in the  elastic-energy linear density,
 via the curvature term $ \frac{1}{2} \,A\,{ \ddot {\vr} }^2 $.
 The mapping  onto  an Euclidian  Quantum Mechanics problem   is no longer as evident as it was 
  before.  
 \subsection{ A very brief review of previous studies on  confined semiflexible polymers.}
       Numerous authors have addressed themselves to the  problem of finding an 
Hamiltonian describing semi-flexible polymers subjected to spatial constraints 
 \cite{Mag,Gom,Burk93,Burk97, Burk01,kier,saito,fried,Yama,helf, mors}. The basic idea 
 was  to formulate the problem in such a way that both the coordinate $ \vr$ and the  ``velocity" $ \vv=\dot{\vr} $ 
appear as  independent  dynamical variables. Various  mathematical techniques have been used to arrive to the Hamiltonian :  
\be
    \widehat{H }= -\frac{1}{2\, A} {\nabla_{\vv}}^2 +{\cal{E}}_0( \vv^2)+\vv \cdot ({\nabla}_{\vr} -\vf) +V(\vr)\,.
\label{Hv}
\ee
The case of the WLC  model is obtained by choosing $ {\cal{E}}_0(\vv^2)  \propto \frac { ( \vv^2-1)^2 }{2 \,\delta b^2 }\, $
and going to the limit $ \delta b\rightarrow 0$. It leads  to the spatially constrained   WLC Hamiltonian: 
\be
{\widehat{H}}_{SCWLC}= {\widehat{H}}_{WLC} +\vt\cdot {\nabla}_{\vr}+V(\vr).
\label{SCWLCH}
\ee
 A group of authors \cite{Mag,Gom,Burk93,Burk97, Burk01,kier}
 have focused  their analysis on the case of semiflexible polymers confined  within a cylindrical tube having a radius much 
smaller than the persistence length. It corresponds to the choice  $ {\cal{E}}_0(\vv^2) =1$ with the running  variable $s$ 
being identified with 
the monomer-coordinate component along the axis of the tube. This can be viewed as  the statistical physics analog of 
the Monge approximation for elastic rods.  This model has been applied recently to  interesting  physical problems:
the first one is   the theoretical study of   the flow of  semiflexible polymers across cylindrical pores \cite{Burk01}, a second one 
is the analysis of the unbinding transition between semiflexible polymers 
and directed polymers acting as one-dimensional attractive  systems. \cite{kier}. 

 The fully three-dimensional WLC Hamiltonian, involving both the coordinate $\vr$ and the ``velocity'' $ \vv=\vt$, appears
 in references \cite{saito,fried,Yama,helf, mors}. One of the most  significant  applications, which bears some 
resemblance with the work presented here, is found in reference \cite{mors}, where the authors analyse a
symmetric interface between two immiscible, semiflexible polymers.

%%%%%%%%%%%%%%%%%%%%%%%%%%%%%%%%%%%
\subsection{ Organization  and Synopsis  of the Paper}
%%%%%%%%%%%%%%%%%%%%%%%%%%%%%%%%%%%%%
The aim of  the present paper is the study of the spatial obstructions which are present in  most 
 micromanipulation experiments, involving  a single dsDNA, immersed in a liquid thermal bath and  
possibly interacting with a single protein. These obstructive effects are expected to be
significant in recent  elongation  experiments,  involving 
molecular segments  with  a cristallographic length $ L$ of about ten persistence lengths \cite{Terence}. 

  We have  not tried  to  obtain the  partition function  of  a  ds-DNA molecule,   subject  to spatial 
constraints,  by solving the  partial differential equation  associated with a Schr\"odinger-like  problem,
written with an  imaginary time equal to $  -i \;s $.
In  Section I, we   rather stay within a discretized version of the WLC Model  
involving $N$ links (or effective monomers)  of length  $b=L/A$. 
% This  is  an usual  
% starting point for the derivation of the  Hamiltonian from the Boltzmann formula.
The partition function $ Z_N$ is obtained by an iteration procedure involving the 
``transfer matrix"  connecting adjacent  monomers \cite{TM}.
 The ``velocities"   $ \vv_n$,  relative to the  $ N $ effective monomers, are  introduced   in the discretized   Boltzmann formula 
via   appropriate Dirac $\delta$-functions.
 The WLC model is obtained by choosing:
$  {\cal{E}}_0(\vv_n^2)= b/( 2 \,\delta b^2 ) \,( \vv_n^2-1)^2 $  and by performing the integrals 
over the moduli  $ v_n= \vert\vv_n \vert$ in the limit $ \delta b/b \ll 1$, so that the tips of $ \vv_n$  are  restricted 
to the unit sphere. After few manipulations,  we arrive to  a recurrence relation  for the partition functions of  
 molecules having their lengths $ L$ which increase by unit of $b$:
 \be
Z_{n+1} ( {\vr}_{n+1},{\vt}_{n+1})=\exp\((-b\,V({\vr}_{n+1})\)) \int\, d^2 \Omega( \vt_n)  
T_{WLC}(\vt_{n+1}\,  \vert\,  \vt_n)\,  Z_n( \vr_{n+1}-b\, {\vt}_{n+1},\vt_n ).
\label{recurZtrvf}
\ee
Here  $T_{WLC}(\vt_{n+1}\,  \vert\,  \vt_n)\,  $ is the transfer matrix relevant for  the unconstrained  WLC model. The above 
 iteration procedure,  which will  be our basic tool for the study of the  
entropic elasticity of dsDNA subject to spatial constraints,   exhibits a suggestive connection with  a  Markovian random  walk in three
dimensions.

The Section II is  devoted to semirealistic   applications  of the above  formalism.
Our testing ground is the study of the entropic elasticity  of a single dsDNA  molecule confined between 
two parallel plates. One molecular end is anchored to one   plate by a biochemical binding. A stretching force,
normal  to the plates, is pulling away the free molecular end from the anchoring plate. As long as we are 
mainly interested in the elongation of the ds-DNA molecule, we can   exploit the invariance under translations 
parallel to the plates and rotations around the stretching force to write a recurrence formula 
involving only  longitudinal variables:
$$
Z_{n+1}( z_{n+1},\theta_{n+1})= \exp\((-b\,V(z_{n+1})  \)) \int_0^{1}\, d\,(cos\theta_n)  
{\cal{T}}_{WLC}({\theta}_{n+1},{\theta }_n, f) Z_n( z_{n+1}-b\,\cos\theta_{n+1},\theta_n),  
$$
where $ z $ and $ \cos \theta$ are respectively the components along the force direction of the  coordinate $ \vr $ and the 
unitary velocity vector $ \vt $.
 
For the sake of simplicity, 
we ignore, in this preliminary  analysis, possible  spatial obstructions associated with the stretching  devices (magnetic 
and optical  tweezers), but, in Section III, we suggest  a practical way to take them into account.
  Two configurations are  studied.

 In the first one, the cristallographic length $ L$ is supposed
to be shorter than the two-plate distance $L_0$, so that the entropic elasticity  is  affected only  by the anchoring plate barrier.
For a relatively short molecule ($ L/A $=12),  we  compute  the  elongation-versus-force curve,
 which is expected  to be  pushed  upward  with   respect to the unconstrained WLC  model predictions,
notably  for the zero-force case. In the case of  relatively high force F, {\it i.e} $\alpha= F\,A/(k_B \,T)=5$, 
the  elongation $ \bra\,z(L)\,\ket$   versus $L$  is very well fitted, within the interval  $ 2 A \leq L \leq  10 A  $,
 by a straight  line with a slope = 0.771 and a constant term $ =1.22\, A$. The slope turns out to be very close to the value  predicted 
by the unconstrained WLC model in the limit $A\ll L$, namely, $0.775$. If the second plate is pushed further away, 
 the predictions of the two models  for  $  \bra\,z(L)\,\ket/L $ will coincide  when $ A\ll L$   up to a $ 0.5 \%$
correction, which could be due to our use of a discretized  version of the  $WLC$ model. 

In the second configuration,  $ L$   is allowed to take values up to  $1.5 \, L_0$,  
with  the same   stretching force  value as above: $\alpha=F\,A/(k_B \,T)=5$.
  A molecule, elongated  according to the unconstrained  WLC model    
cannot fit within  the plates when  its  cristallographic length $ L$ is
 larger than  the  critical value   $L_*= 1.29 L_0$, associated with the elongation  $  \bra\,z(L_*)\,\ket=L_0.$ 
We have studied the evolution of   the terminal monomer  statistics  when $L$ varies within the interval: $ 0.6 \, L_0 \leq L\leq 1.5 \,L_0$,
using as representative quantities the elongation  derivative    $ d\; \bra\,z(L)\,\ket/ dL $ and  the  mean square 
of the free-end fluctuations
  along the  stretching force direction, {\it i.e.}  $  \Delta \, z^2 (L)= \langle ( z(L)-\langle z(L) \rangle)^2 \rangle $.  When $ L\le L_0$  these 
two quantities follow rather closely the predictions of the unconstrained WLC model $  $  but they both start 
 a rather sharp decrease  when $ L >L_0$. The elongation $  \bra\,z(L)\,\ket $ is no longer an extensive 
 physical quantity an goes slowly to $L_0$. As to the fluctuations $ \Delta \, z^2 (L) $, they are reduced by a factor 10  with respect to 
the unconstrained WLC model  prediction.

In the case   $L= L_{max}= 1.5 \,L_0 $, we give a semi-quantitative analysis of  the  internal monomers statistics, when
 the monomer   number  varies within the  interval $ L_*/b  < n < L_{max}/b$.
 The characteristic feature  of our model lies in the fact 
that {\it all the monomers are confined } between the two plates. We call  this type of spatial
 constraint   { \it  an internal confinement} (IC)  in  contrast with   the external confinement (EC) describing 
a situation where the confining $
V(z) $ is {\it  acting  only upon the terminal monomer}.   A  rather good physical
understanding of what is going on
 near  the repulsive barrier has been obtained within   a Gaussian model,  supplemented by a square-well potential
 with a depth $\gg k_B \, T$. Choosing a configuration  with the same value of $ L_* $ as
 before,  we find, without too much surprise, that in the IC model
the terminal segment  involving   $ 20 \,\% $ of the internal monomers flattens  
against the  repulsive barrier,  while in the EC model all the
monomers, except the terminal one, can  move  rather freely across the repulsive  barrier.
 
In Section III, we suggest possible   
applications  or extensions of the work  presented in the present paper; let us mention the two of them
which concern  directly  the  micro-manipulation expriments:
 \begin{itemize} 
 \item  We propose a simple procedure involving a   two-plate confinement  model  which may lead  to an
estimate of the spatial obstruction effects  associated with the magnetic tweezer.
   \item It is relatively  straightforward  to generalize the  spatially constrained  WLC model,  within its 
  transfer matrix formulation, to the RLC model \cite{BouMez98,moroz,BouMez00}, which incorporates both bending and
twisting  rigidities. The  anchoring  barrier effects are expected to be significant when the reduced supercoiling parameter $\sigma
$ is above the threshold - at fixed force - for the creation of plectonem configurations. The stretching potential 
energy vanishes for such structures, allowing them to wander irrespective of the sign of $ \vf  \cdot  \vt $.
\end{itemize}

%%%%%%%%%%%%%%%%%
The Appendix is devoted to a comparison  of two possible  approaches to the elastic entropy elasticity of dsDNA subject to 
spatial constraints. The first one, used in  the present paper,  is the transfer matrix method which leads to an iterative construction  of
the partition function, within a discretized version of the constrained  WLC model.  The second  approach
 \cite{Mag,Gom,Burk93,Burk97,Burk01,kier,saito,fried,Yama,helf, mors}
 is based upon the solution  of a Schr\" odinger-like equation,
 written with an imaginary time variable $-i \, s$   involving the Hamiltonian given by eq. (\ref{SCWLCH}).  
To achieve our purpose,  we have found convenient to use  the auxiliary variable  method  in order to eliminate 
 the second order derivative in the  elastic energy density given by eq. (\ref{elasdens}).  It is, then,  a rather straightforward
affair  to obtain  the  transfer operator 
$ \widehat{T}$- associated  with the transfer matrix-, from which 
one   derives  the    Hamiltonian  $ \widehat{H}$  by taking  the continuous limit $ b \rightarrow 0 $. 
 Making use of the leeway inherent to  any
discretization  procedure,  one can   obtain a symmetric version of $ \widehat{T}$,  which coincides 
with the exact evolution operator $\exp-b \, \widehat{H}$,  up to corrections of the order of $ b^3$.

It is not too difficult to  elucidate the physical interpretation of the auxiliary variable $\vu$: we 
have proved  that the conjugate momentum $ \vp_u $ appearing in $ \widehat{H}$ is just the ``velocity"  $ \vv$.
Performing an  appropriate change of basis, one obtains immediately  an expression of    $ \widehat{H}$,  identical to the  
r.h.s. of eq.(\ref{Hv}). With  the proper choice of  ${\cal{E}}_0(\dot{\vr}^2 )$, one then arrives  to the spatially constrained WLC
Hamiltonian, $ \widehat{H}_{SCWLC}$,  given by  eq.(\ref{SCWLCH})
\cite{Mag,Gom,Burk93,Burk97,Burk01,kier,saito,fried,Yama,helf, mors}.
Finally, as an internal check, we derive a symmetric transfer matrix  from  the transfer operator $ \widehat{T}$, associated
with the Hamiltonian $ \widehat{H}_{SCWLC}$. By making suitable approximations, we recover  the  transfer matrix,
physically more transparent, which is  derived directly from the Boltzmann formula in Section I.

  %%%%%%%%%%%%%%
\section{A Transfer Matrix  approach To   Stretched ds-DNA  subject to Spatial Constraints }
  We wish, first, to consider  the statistical properties of a class of polymer models    described  by the linear elastic-energy density ${\cal{E}}(s)$
depending upon the monomer coordinate $ \vr(s) $, together  with its first and second order derivatives, $\dot{\vr} = \frac{d\vr }{ d s}$ and $
\ddot{\vr} = \frac{d^2}{ds^2}(\vr)$. The variable  $s$ with  $0 \leq s \leq L  $ results  from a coarse graining  of  the molecular  chain,  having
a cristallographic length $L$. Note that  $s$ 
does not coincide necessarily  with the arc-length  $ s$  of a  rectifiable curve. 

 The  partition function  for   fixed free-ends  polymers is given by a functional  integral involving the 
standard  Boltzmann  stastitical weights      :
\bea
Z = \int {\cal{D}} \ [ \,\vr \ ] \exp{ \left( -\int_0^L {\cal{E}}(s) ds \))} \,,\\
 {\cal{E}}(s) = {\cal{E}}_0(\dot{\vr}^2 ) + \frac{1}{2} \,A\,{ \ddot {\vr} }^2  - \vf \, \dot {\vr}  + V(\vr)\, ,
\label{elasdens}
\eea 
where the functional  integration   goes over all the paths joining the free ends 
of the chain of fixed coordinates $ { \vr(0)} $ and $ { \vr(L)} $.

We stress that, here, the potential $ V(\vr)
$,  simulating the  spatial constraints,
  {\it is clearly acting upon all the monomers of the molecular  chain}, contrary to confining  models where 
 the potential  $ V(\vr )$ is acting, only, upon the terminal monomer, {\it e.g. via} an attached bead,  
  In this  latter case, constraints   can be trivially implemented by adding  in the elastic density    (\ref{elasdens}) the ``velocity"
dependent  contribution $ \dot{\vr}\cdot \nabla_{ \vr} \,V( \vr)$.

To compute the partition function  we have to  resort to a discretization of the variable  $s$: $ s_n =n\, b$.  The molecular
chain  is then   represented  by $ N $ elementary links  or effective monomers  with $ N=L/b$.
Assuming that   the effective monomer  length $b $ is  much smaller than the persistence  length   $A$, 
 we can write:
\be
\dot{\vr_n } =\frac{ \vr_n - \vr_{n-1} }{b}, \hspace{15mm} \ddot{\vr_n } = \frac{\vr_n - 2\vr_{n-1}+ \vr_{n-2} }{b^2} \,.
\label{discmoner}
\ee 
It is then convenient to  introduce the discretized form of the elastic-energy density ${\cal{E}}_{disc}(n)$  obtained by plugging
the formulas (\ref{discmoner}) into the right hand side (r.h.s.) of equation (\ref{elasdens}):
\be
   {\cal{E}}_{disc}(n)= {\cal{E}}_0 \(( ( \frac{ \vr_n - \vr_{n-1} }{b} )^2 \))+
 \frac{1}{2} \,A\, (  \frac{\vr_n - 2 \,  \vr_{n-1}+ \vr_{n-2} }{b^2}    )^2 -\vf \, \cdot  ( \frac{ \vr_n - \vr_{n-1} }{b}) +
 V(\vr_n) .
\ee
The partition  function  is then given  as an   integral  of a product of    $ N-1$  monomer-coordinate-$\vr_n$ functions:  
\be
 Z =\int \prod _{n=1}^{N-1} d^3\vr_n  \, \{ \prod _{n=2}^{N}  \exp \((- b  {\cal{E}}_{disc}(n) \)) \}\, Z_{in}( \vr_1,  \vr_0) ,
\label{intfuncZ}
\ee
where the monomers coordinates $\vr_0$ and $\vr_N$  are kept fixed.
The starting partition function  is taken to be :  
\be
 Z_{in}( \vr_1,  \vr_0)=  \exp \(( - b   {\cal{E}}_0 \(( ( \frac{ \vr_1 - \vr_{0} }{b} )^2 \)) -b V(\vr_1) \)) \,. 
\ee
From the above expression of the partition function one  could write down easily a transfer matrix connecting 
the probability  distributions  relative   to  pairs  of adjacent monomers:
\be
Z_{n+1}( \vr_{n+1},  \vr_n)=  \int  d^3\vr_{n-1} d^3\vr_{n-2} \exp \((- b \,  {\cal{E}}_{disc}(n+1)  -b \,  {\cal{E}}_{disc}(n) \))
Z_{n-1}( \vr_{n-1},\vr_{n-2}) .
\ee

This approach turns out to be  rather awkward in the particular case of the WLC model but may be useful for other models.
The standard trick is to introduce  the ``velocities "  ${\vv}_n=\dot\vr_n   =( \vr_n - \vr_{n-1})/b$ as new variables in
the discretized functional  integral (\ref{intfuncZ})  via the trivial identity: 
$$  1=   \int  d^3 \vv_{n} \, \delta^3( {\vv}_n - \frac{ \vr_n - \vr_{n-1} }{b} ) .$$
   Making some obvious manipulations,  one gets a  new expression  for the partition function as an  integral over  paths  joining 
 two elements in the  6-dimension space obtained  by taking 
 the direct sum  of the coordinate $ \vr$  and velocity $ \vv$ spaces:
  \bea
Z_N(\vr_N,\vv_N) &=&\int \prod _{n=1}^{N-1} d^3\vr_n  \,  d^3\vv_n \, \{  \prod _{n=2}^{N} \,  
\delta^3( {\vv}_n - \frac{ \vr_n - \vr_{n-1} }{b})    \nonumber \\  
& &\exp-b \(( {\cal{E}}_0  ( \vv_n ^2 )+\frac{1}{2} \,A\, ( \frac{ \vv_n -\vv_{n-1}} {b}   )^2 - \vf \cdot \vv_n+V(\vr_n) \)) \}\,
Z_1(\vr_1,\vv_1) .
\label{Zrv}
 \eea 
For practical reasons, the coordinate $\vr_0$ will be assumed from now on  to be a random variable with  the probability distribution $P(\vr_0)$.
Then, $ Z_1(\vr_1,\vv_1)$ is given by:
\be
Z_1(\vr_1,\vv_1)=\exp-b \(( {\cal{E}}_0 ( \vv_1 ^2 ) - \vf \cdot \vv_1 +V(\vr_1) \))  P_0( \vr_1-b\,\vv_1).
\ee
If one performs explicitly the integration upon the $ \vv_n $ variables in the r.h.s of equation  (\ref{Zrv}), one  readily recovers 
 the r.h.s of equation (\ref{intfuncZ}) with $ Z_{in}( \vr_1,  \vr_0) $  replaced by its average over $\vr_0$.
Only nearest-neighbour monomers with given coordinate  and velocity are now 
 connected.  It is then  possible to write down a recurrence relation between 
  adjacent intermediate partition functions  $ Z_n ( \vr_n, \vv_n ) $  relative to  chains having 
a  cristallographic length $ s_n= b\, n $: 
\be
Z_{n+1} ( \vr_{n+1}, \vv_{n+1} ) = \int \int  d^3\,\vr_n \,d^3\,\vv_n \,T( \vr_{n+1},  \vv_{n+1}\,  \vert\, \vr_n,\vv_n)\,
Z_n( \vr_n, \vv_n) . 
\ee
 The    transfer  matrix $  T( \vr_{n+1},  \vv_{n+1}\,\vert \, \vr_n,\vu_n) $   is easily  read 
 off   from  the r.h.s of equation (\ref{Zrv}):
\bea
T( \vr_{n+1}, \vv_{n+1}\,  \vert\, \vr_n,\vv_n) &= & \delta^3( {\vv}_{n+1} - \frac{\vr_{n+1} - \vr_{n} }{b})
 \, \exp- b\((    \,V(\vr_{n+1} ) \)) \times \nonumber \\ 
   & & \exp-b \(( {\cal{E}}_0  ( \vv_{n+1}^2 )+\frac{1}{2} \,A\, (  \frac{\vv_{n+1} -\vv_n}{b}    )^2 - \vf \cdot \vv_{n+1}\)) .
 \eea
Performing the integration upon  $\vr_n$,    we arrive  to the   the final form of the  recurrence relation:
\bea
Z_{n+1} ( \vr_{n+1}, \vv_{n+1} )&=&\exp-b \((  {\cal{E}}_0 (  \vv_{n+1} )^2 +V(\vr_{n+1}) \)) \,  
   \int  d^3\,\vv_n   \times    \nonumber \\
& & 
  \exp- \(( \frac{1}{2 b} \,A\, (\vv_{n+1} -\vv_n)^2 -b\, \vf \cdot \vv_{n+1} \)) \, Z_n( \vr_{n+1}- b \,
\vv_{n+1}\, ,
\vv_n) . 
\label{reccurel}
\eea
In order to apply the above formula   to the WLC  model with spatial constraints,   we have  to choose an adequate
form for the function   ${\cal{E}}_0(\vv^2)$.  The solution   is rather simple: 
one introduces the small length $ \delta b$ such that 
$ \delta\,b/b \ll 1$ and one takes  for  $ {\cal{E}}_0(\vv^2) $ the following expression:
  \be
{\cal{E}}_0(\vv_n^2)= b\frac { ( \vv_n^2-1)^2 }{2 \,\delta b^2  }\,.
\ee
Indeed, with this choice, we write the volume element in the velocity  space as: $ \int  d^3\,\vv_n = v_n^2\, dv_n  \, d^2 \Omega( \vt_n)$, 
where
$d^2 \Omega(\vt_n)$  is the infinitesimal solid angle around the unit vector $ \vt_n $ taken along $ \vv_n $, written as $ v_n \, \vt_n $.  
Then, in the  limit  $ \delta\,b/b  \rightarrow \, 0$ ,  $ \exp-b\(( {\cal{E}}_0  (\vv_n ^2)\))$ 
  reduces, up   to a numerical constant, to the $\delta$-function: $\delta (v_n-1) $.  As  a consequence,   integrating over $v_n$
 within the same limit, leads to an expression of the r.h.s  of  equation ($\ref{Zrv}$) readily obtained by making the 
replacement: $$ d^3\,\vv_n  \rightarrow  d^2 \Omega( \vt_n)\,,  \hspace{15mm} \vv_n \rightarrow \vt_n. $$
  The transfer matrix describing stretched dsDNA  within the WLC model with spatial constraints is then 
simply obtained from eq.(\ref{reccurel}) using the same  replacement rules:
\bea
Z_{n+1} ( \vr_{n+1}, \vt_{n+1} )&=& \exp-b \((V(\vr_{n+1}) \)) \,  \int d^2 \Omega( \vt_n) \times 
   \nonumber \\
& &     \exp-\(( \frac{A}{2}(\vt_{n+1} -\vt_n)^2 - \vf\cdot (\vt_{n+1}+\vt_n)/2\))  
\, Z_n( \vr_{n+1}- b \, \vt_{n+1}\, , \vt_n) . 
\label{reccurelWLC}
\eea
To make easier the comparison with the  standard WLC model, we have replaced in the above equation $ \vf \cdot \vt_{n+1}$ 
by $\vf\cdot (\vt_{n+1}+\vt_n)/2$. Such a modification  amounts to the replacement of a rectagular discretized integration 
by a trapezoidal one.
What appears now in  the r.h.s of the recurrence relation is  the transfer matrix relative to the standard WLC model:
\be
T_{WLC}(\vt_{n+1}\,  \vert\,  \vt_n) = 
\exp\((- \frac{ A \, (\vt_{n+1}-\vt_n)^2} {2 \, b}+
\frac{b}{2} \,\vf \cdot (\vt_{n+1}+\vt_n)\,\)) \,.
\label{trmatWLC}
\ee
In the Appendix, we are  going to show, within a more general  context, that $T_{WLC}(\vt_{n+1}\,  \vert\,  \vt_n)$  
coincides,  up to corrections of the order of $ (b/A)^3$,
 with  the exact transfer operator $ \exp\((-b\,{\widehat{H}}_{WLC} \))$   with 
${\widehat{H}}_{WLC} = -\frac{1}{2\, A} {\nabla}_{\vt}^2 - \vf \cdot \vt $.

We are, now, ready to give the partition-function recurrence relation  which will  be our basic tool for the study of the  
entropic elasticity of dsDNA subject to spatial constraints:
\be
Z_{n+1} ( {\vr}_{n+1},{\vt}_{n+1})=\exp\((-b\,V({\vr}_{n+1})\)) \int\, d^2 \Omega( \vt_n)  
T_{WLC}(\vt_{n+1}\,  \vert\,  \vt_n)\,  Z_n( \vr_{n+1}-b\, {\vt}_{n+1},\vt_n )\,.
\label{recurZtrvf}
 \ee

The iterative construction  of the partition function  $Z_n(\vr_n,\vt_n)$ has a transparent physical meaning. Indeed, it 
 has    a suggestive  interpretation in terms  of  a Markovian random walk model in three dimensions:
the $(n+1)^{th}$ step is given by $ b\, {\vt}_{n+1} $; its length  is $b$ and its direction, defined by the unitary  vector $ {\vt}_{n+1}$,
 is correlated to that of the previous step   $\vt_n $  through   the correlation function  
$ C({\vt}_{n+1},\vt_n ) \propto T_{WLC}(\vt_{n+1}\,  \vert\,  \vt_n) $.
Within our unit conventions,  the ``potential"  $ V(\vr)$   stands for a potential  energy by unit molecular length, written in 
thermal units,  so that
$ \Delta W_{n+1} =   k_B \, T\,  b\,V({\vr}_{n+1})$  represents the amount of energy   to be exchanged with
 the thermal  bath 
by  performing the  $ (n+1)^{th}$ step. The exponential   in front of the r.h.s of
  equation  (\ref{recurZtrvf}) is just the associated  Boltzmann factor.

%%%%%%%%%
\section {  The stretching  of a ds-DNA  molecule confined  between two parallel plates. }
%%%%%%%%%%%%%
In this section, we apply our iterative version  eq.(\ref{recurZtrvf}) of the spatially  constrained WLC  model  to the confinement
of a single dsDNA molecule between two parallel plates - called hereafter Plate 1 and Plate 2 - separated by a distance $L_0$. One
molecular end is anchored upon Plate 1. The other end is attached, for instance, to a magnetic bead pulled by  a magnetic tweezer. We
do not  pretend, here,  to describe  a fully realistic situation  since,  for the sake of simplicity,
 we   ignore  the spatial constraints
associated with  the finite dimensions of the magnetic bead.  In  Section III of this paper, 
 we shall suggest  a self-consistent procedure, within
the two-plate model, which will lead to an   estimate of the bead obstruction effect. 
  The stretching force is assumed to  be normal to the two plates and is pulling away the 
molecular end   from the  anchoring plate.  The dimensions of the plates normal to the force, measured from 
the anchoring point, are assumed to be  much larger than  $L_0$ and will be considered as infinite. 
Two physical  configurations  have been considered: 
\begin{enumerate}
  \item  $ L_0$ is supposed to be larger than the cristallographic length $L$. (In
our explicit  computations we have taken  $ L_0= 40~A $ and $L= 12~A$.)  It means that even for large stretching 
forces the molecular end stays far away from Plate 2. This  will allow us to  study  the deviations of the elongation  from
the WLC model prediction   due  to the   presence of the anchoring  plate  (Plate 1), provided we ignore the  spatial obstruction of 
the pulling device.
 \item  The cristallographic  length $L$  is larger than $L_0$ so that, for  large  forces, the  fully stretched molecule does not fit  inside
the plates. We have studied,  within the  WLC model,   how  the molecule adapts itself to a confined situation 
  where the elongation  predicted by the unconstrained WLC Model  reaches 
 values  significantly larger than   $L_0$.
\end{enumerate}
%%%%%%%%%%%%%%%%%%%%%
\subsection{Basic formulae for  the two-plate confinement configurations.}
Our confinement configurations, defined  by the two confining plates and the stretching force, are invariant, 
first, under  rotations around  the  $z$-axis defined by the anchoring
point and the stretching force direction, second, under the translations within the  $x,y$ plane. We shall  consider a
molecular ensemble  where the anchoring monomer is uniformely distributed over the $x,y$ plane. 
The iteration procedure  can then  be organized in such a way that at each step 
the partition function  $ Z_n(\vr_ n\, ,\vt_n) $ depends  only upon the two {\it independent } 
variables:  $ z_n $  and $ \cos {\theta_n }$,   which are respectively the components  upon the $z$-axis  
of the  monomer coordinate  $\vr_ n$ and   the tangent vector $\vt_n$. 
If this condition is satisfied at the step $n$,   $Z_{n+1}( z_{n+1},\theta_{n+1})$  is then easily
obtained  from an azymuthal average around the $z$-axis of the two sides of eq.(\ref{recurZtrvf}):

\bea 
Z_{n+1}( z_{n+1},\theta_{n+1})&=& \exp\((-b\,V(z_{n+1})  \))
\int_0^{1} \hspace{-2mm}d\,(cos\theta_n) \, 
{\cal{T}}_{WLC}({\theta}_{n+1},{\theta }_n, f)\, \times \nonumber \\
& & Z_n( z_{n+1}-b\,\cos\theta_{n+1},\theta_n)  
\label{recurZ2plates}
\eea
The  azymuthal-averaged WLC transfer  matrix  can be obtained by taking the zero torque limit  of  the corresponding 
 expression   given  in  ref.\cite{BouMez00} for   the  supercoiled DNA case:
   \bea
{\cal{T}}_{WLC}({\theta }_1,{\theta }_2, f) & = &\exp-\lbrace \frac{A}{b}\, \left( 1  -
                                  \cos {\theta }_1\, \cos{\theta }_2  \right) 
   +\frac{b\,f}{2} \left( \cos {\theta }_1 + \cos {\theta }_2 \right)  \rbrace \nonumber\\
& & \times  \,I_0 ( \frac{ A \sin{\theta }_1\, \sin{\theta }_2 }{b}   )
\label{TWLC }\,.
\eea 
%%%%%%%  FIG1%%%%%%
\begin{figure}
\vspace{ 10mm}
%\centerline{\epsfxsize=100mm\epsfbox{FigSC1.eps}}
\centerline{\epsfxsize=120mm\epsfbox{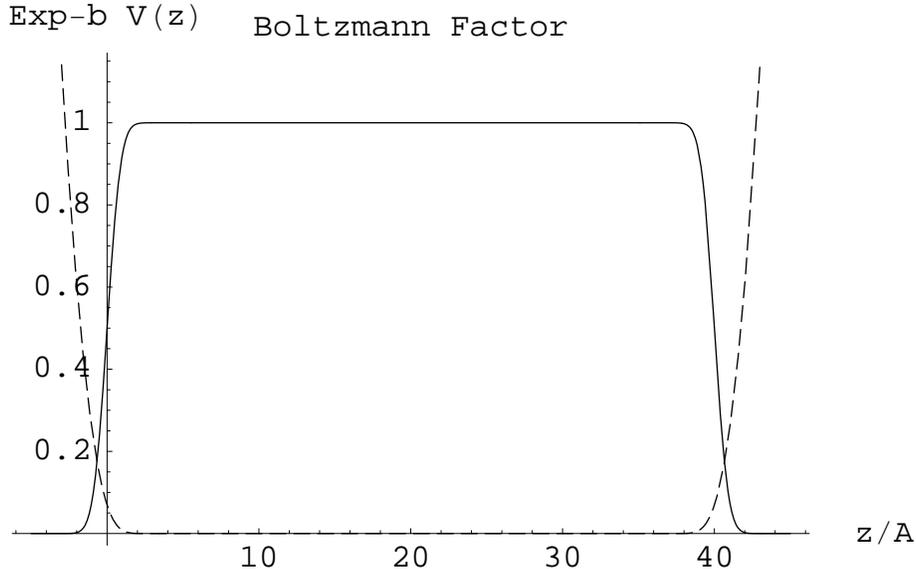}}
\caption{ \small The two-plate confining  potential $V(z)$  (dotted line)
 with the  associated Boltzmann factor $\exp -b\, V(z) $ (full line). We have taken as  elementary link length  $ b=A/10 $ .
 The   Boltzmann factor is given explicitly by the product of two rounded-off step functions
  $\Theta( z,\Delta z) \, \Theta(L_0- z,\Delta z)$   
with a smoothing length $\Delta z= A$ and a two-plate distance $L_0 = 40 A$. } 
\label{fig1}
\end{figure}
In order to build   a potential  $ V(z) $ simulating the two confining plates and  satisfying a
smoothness condition at the persistence-length scale, it is convenient  to introduce the rounded-off step function:
\be
\Theta( z,\Delta z)=\frac{1}{2} + \frac{1}{2} {\rm erf}( z/\Delta z) \,, 
\ee
where $ {\rm erf}(x) $  is the  ``error" function : $ \frac{2}{\sqrt{\pi }} {\int}_0^x \exp (- \,t^2) \,d\,t $ and $\Delta z$
the smoothing length  assumed to be $\sim \,A $.  Instead of the potential  $ V(z)$,  it is more convenient to   write directly the
Boltzman factor :
\be
{\rm  Boltz}( z,L_0,\Delta z)= \exp\(( -b\,V(z) \))=\Theta( z,\Delta z) \, \Theta(L_0- z,\Delta z)\,.
\ee
Its variation with respect to the reduced variable $z/A$ is represented by the solid curve of Figure 1 for $ \Delta z=A$. 
The dotted curve gives the variation of the external potential  associated 
with an elementary link~$ b\,V(z_n) $.

The next point  to be  specified  is the inital condition of our  recurrence procedure.  A detailed analysis  of the anchoring 
mechanism at the microscopic level being   beyond   the scope of this paper, we have adopted a   prescription  which
incorporates in a simple way  some features of the actual physical situation. 
We choose as  the molecular chain origin  ($n=0$)
 the end of the  initial  strand of length $ \sim A $ sticking out from the anchoring plate. The
initial  partition function $ Z_0( z_0,\theta_0) $  is written as follows :
\be
 Z_0( z_0,\theta_0)=\frac{1}{2\,\sqrt{2\,\pi} \,\sigma}\exp\(( -\frac{1}{2} ( z_0/\sigma)^2  \)) .
\ee
A look at the potential $V(z $)  on Fig. 1 shows  that $ z_0$ can take negative as well as positive values within the range
$ (-A,A )$,  so that a centered Gaussian curve looks quite  realistic if $ \sigma \sim A$. Similarly the tangent-vector  
 $z$-axis projection, $\cos \theta_0 $, lies within the interval $ (-1,1 )$, so that a   $\cos \theta_0 $  uniform distribution  seems 
to be an acceptable guess. We note, finally, that our choice of  $Z_0( z_0,\theta_0)$ satisfies the smoothness condition assumed
in the previous section. The $n=1$ partition  function $ Z_1( z_1,\theta_1) $ is then readily obtained by plugging $Z_0(
z_0,\theta_0)$ in the  r.h.s. of the recurrence relation (\ref{recurZ2plates}).

We would like  to give, now, a few indications about the numerical methods we have used 
 to run  the transfer matrix iteration process. 

In  order to perform the  integral over $ {\theta }_n$,   we use the following  
discretization procedure. We divide the variation interval $ 0 \leq {\theta }_{n} 
 \leq  \pi $ into 
$ n_s $  segments $ { (s-1)\pi \over n_s } \leq {\theta }_{n} \leq  { s\,\pi 
\over n_s } $.
The integral over  each segment is done with   the standard Gauss method  
involving
$ n_g $ abscissae and $ n_g $ attached weights. The integral over the full $ \theta_n$
   interval is then  approximately given by a discrete weighted  sum over $ d =n_s \, 
n_g $ points:
$$
 \int_0^{\pi}  f( {\theta }_{n})\sin {\theta}_{n} d{\theta}_{n}=
\sum_{i=1}^ {d} w_i \,\sin{\theta}_i f( {\theta}_i ) \, .
$$

 The  transfer-matrix iteration involves a $z$ variable translation $ z_n\rightarrow 
z_{n+1}- b \cos \theta_{n+1}$  to be performed  upon $Z _n(z_n,\theta_n)$ for each value $ \theta_n= \theta_i $. This is 
achieved by  building  at each step  the  interpolating function  $Z_{n ,int}(z, \theta_i ) $  associated with  the array: 
$ \{ z_l  =z_{min}+l\, b ,Z_n(z_l,\theta_i)   \}$  with  $1\leq  l\leq   n_{max }$.  An appropriate choice  of   $n_{max } $    
 leads  to a physically relevant sample of the    monomer  coordinates throughout the iteration process. 

For more extensive computations than those presented in this paper, one should  consider a  potentially 
more efficient  method involving  the  Fast Fourier Transform  (FFT)  algorithm. It  is based upon the remark 
that a variable translation performed upon a given function reduces to a phase shift  upon  its Fourier  transform: 
$ \widetilde { Z}( p, \theta_i) \rightarrow \exp\(( -\, i b \cos \theta_i  \)) \, \widetilde { Z}( p, \theta_i). $  An inverse 
FFT will then  be required to perform  the multiplication by the Boltzmann factor. 
%%%%%%%%%%%%%%%%%%%%%%%%
\subsection{Modification of the  elongation-versus-force   curve induced by the anchoring  plate barrier.  }

We would like to present, here, the results of a numerical simulation based upon the  iteration process
built up from the recurrence relation given in eq.(\ref{recurZ2plates}). Our aim was to study  the barrier effect of the 
anchoring plate (Plate 1.) upon the elongation-versus-force  curve.  This corresponds to the configuration~1,
introduced previously, where the distance  between the plates $L_0$ is larger
 than the cristallographic length  $ L $. The values used in our simulation are $ L_0=40~A $ and $ L=12~A$. It is, then, clear 
that  Plate-2  plays no role since  we  have decided to ignore 
 the eventual  presence of a  magnetic bead  attached to the molecule free end.  
  At each step $n$  we store the normalized $z_n$-probability distribution, $P_n(z_n,\alpha)$,  where 
$ \alpha= F\, A/(k_B \, T) $ is the  reduced force parameter. We have  plotted in  Figure 2
 the relative molecular elongation   $\bra\, z(L) \,\ket/ L = \int dz_N \, z_N \, P_N (z_N, \alpha) /L $  versus  $ \alpha $, 
where $ N= L/b$.  The dotted curve represents  the prediction of the  standard WLC  model which 
assumes that  the anchoring device  reduces to a point. 
%%%%%%%  FIG2%%%%%%
\begin{figure}
\centerline{\epsfxsize=80mm\epsfbox{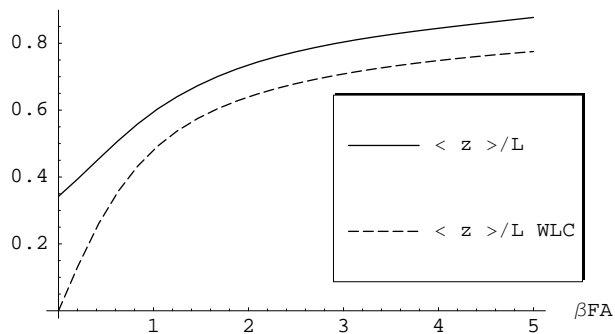}}
%\centerline{\epsfxsize=100mm\epsfbox{FigSC2.eps}}
%\vspace{ -30mm}
\caption{ \small Illustration of the effects of the  anchoring plate barrier upon the elongation-versus-force curve.
The upper curve (full line) gives the prediction of  the spatially-constained  WLC model when $ L/A=12$.  The other plate is  lying
at a distance $ L_0=10  \,L/3$ and, as a consequence, has  no effect upon the elongation.The fact that the   elongation 
curve is pushed upward  with respect to the unconstrained WLC prediction(dotted curve)   has a simple qualitative explanation:  because of
the anchoring plate  barrier the $z<0$ half space is  not accessible to the nucleotides, whatever the stretching force.}
\label{fig2}
\end{figure}
 
The full curve  corresponds to the elongation-versus-force  obtained in our simulation  within a spatially 
 constrained WLC model; it exhibits clearly the  anchoring-plate barrier   effect  which leads to a relative elongation of  $
34\%$ in  the zero force limit. For higher  forces,  the two curves become approximately parallel with an offset of about
$10\%$. It is possible to  get  from our simulation an expansion  in powers of   $\sqrt{ A/L} $ at fixed  force,  giving 
 the upward  elongation  displacement  $\Delta_{bar} \langle  z /L\rangle $ induced by the anchoring plate barrier.  
As an illustration, we give the result for two typical forces corresponding to $ \alpha=0 $  and  $ \alpha=1 $. 
We found that the  second-order  expansion,
      $\Delta_{bar} \langle  z /L\rangle =a_1(\alpha )  \sqrt{\frac{A}{L}} + a_2(\alpha)   \frac{A}{L}$,
   gives a very  good fit to our  simulation data within the range: $ 5\, A \leq   L \leq 30 \, A$.  
For $ \alpha=0 $ we get: $ a_1(0)=1.00 \, , a_2(0)=0.65 $. 
The large value of $ a_1(0)$   can be interpreted  in terms  of an hemispheric molecular cluster around  the
anchoring  point with a radius growing like $ \sqrt{N} $. The results look rather different when  $\alpha=1 $: 
 $ a_1(1)=-.06\, ,  a_2(1)= 1.57 $. The strong decrease
      of  the $\sqrt{\frac{A}{L}}$  coefficient - confirmed by the  $ \alpha=2 $ results  - seems to suggest that the
anchoring  plate barrier affects only a dsDNA segment  having a fixed length $ \sim A$ when   $ \alpha \geq 1 $.

%%%%%%%%%%%%%%%% FIG.3%%%%%%%%
\begin{figure}
  \centerline{\epsfxsize=100mm\epsfbox{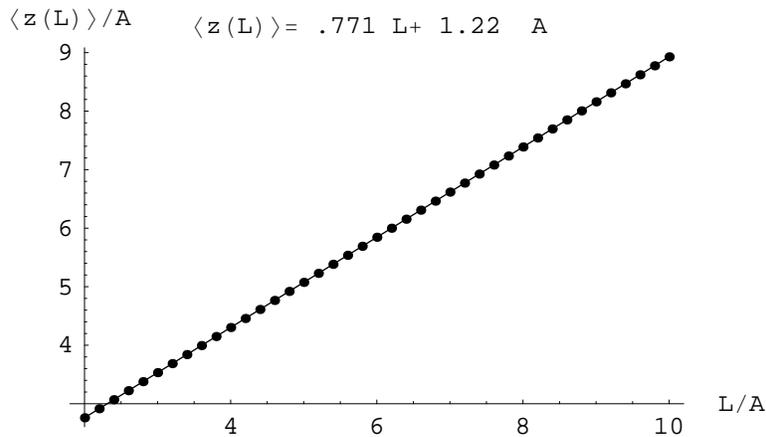}}
\caption{
%On this figure we present a 
Linear fit to a set of    $ \langle\,  z (L)\, \rangle $    values,  coming out from
our iteration procedure  in the case $\alpha=5$, when $ L$  increases from $2\, A$  to $ 10 \,A$. The slope given by
the fit, 0.771 is in good agreement with the prediction of the standard WLC model, 0775, valid  in the  limit $ A/L \ll 1$.}
\end{figure}

When one looks at the curves of Fig.2, one may wonder if  they are going to meet in  the   limit of high force, as they should,  since 
$ \langle\,  z (L)\rangle /L $ goes to one for  the  two cases.  Before trying to push the present computation to larger values of $\alpha$,
one has to keep in mind two things, one, the dotted curve corresponds to the limit $ A/L \ll 1$  while we are working with $ A/L \simeq 0.1$,
second, the dotted curve  goes rather slowly to the limit $ \alpha \to \infty$: indeed, 
$ \langle\,  z (L)\rangle /L $ =0.95 for  $\alpha \simeq 100$. Our computation will simply not work  under these two extreme conditions. 
There is fortunately another way  to make a meaningful comparison of the two models: it is to look at the
derivatives  $\frac{ d \bra\, z (L)\,\ket}{d L}$ instead of $ \langle\,  z (L)\rangle /L $.  In the unconstrained WLC model  computations 
these two quantities are  identical but this is not in general  true in presence  of spatial constraints. We have plotted in Fig. 3  the values
  $ \langle\,  z (L)\, \rangle $     obtained by our iteration procedure in the case  $  \alpha=5$. The results   correspond 
to  the big dots   and cover the range of $ L$ values: $ 2 \,A \leq L \leq 10 \,A $. We have performed a linear fit to the data,
$ \langle\,  z (L)\rangle _{fit}=0.771 L+ 1.22  A  $, which, as it is apparent on Fig. 3, works  beautifully. The fitted slope, 0.771,
is very close  to that given by  the unconstrained   WLC model, namely, 0.775. We also note that the constant   term  1.22  A 
 has  the same order of magnitude as the one obtained previously for $ \alpha=1. $ This strongly suggests that, if the Plate-2 is
pushed to infinity,     the two models will give  identical predictions,  for the elongation $ \langle\,  z (L)\rangle /L $    in the limit $ A\ll L$.
 We have ignored  the   $ 0.5 \,  \%$ 
 difference, which  may  be  attributed to our use of  a discretized model wih $b/A=0.1 $.

%%%%%%%%%%%%%%%%%%%%%%%%%%%%%%%%%%%%%%%
\subsection{Statistical   behaviour of  stretched  molecules having   cristallographic lengths   
larger than the two-plate distance.}    
%%%%%%%%%%%%%%%%%%
%%%%%%%  FIG4%%%%e
\begin{figure}
\centerline{\epsfxsize=120mm\epsfbox{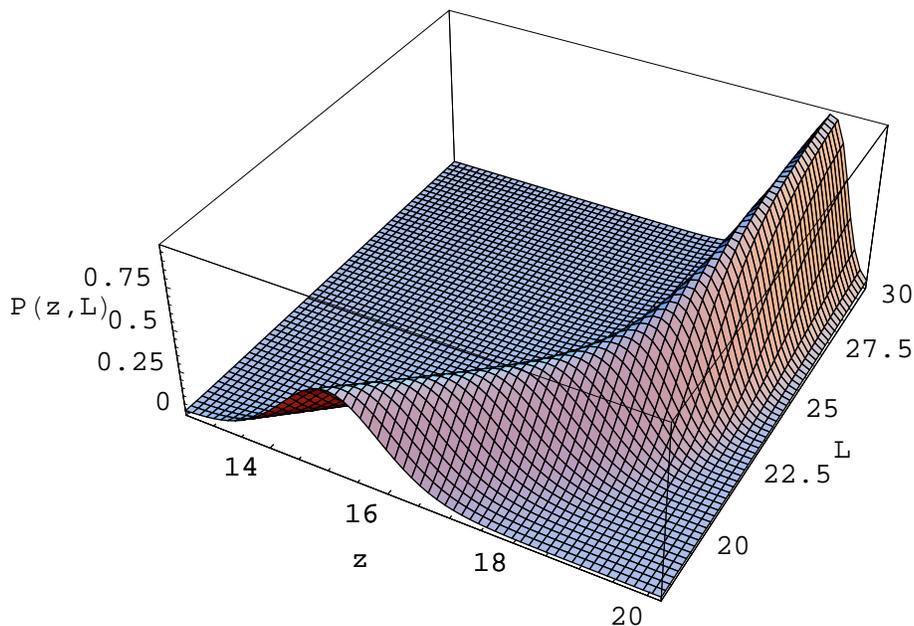}}
%\centerline{\epsfxsize=100mm\epsfbox{FigSC4.eps}}
%\vspace{ -30mm}
\caption{ \small   3-D plot illustrating the probability distribution of the coordinate $z$
of the molecular end, $P(z,L)$,  when the cristallographic length $L$ varies from  $0.9 \,L_0$ to $ 1.5\,   L_0$  with $L_0= 20~A$. 
 In our iterative computation, 
we have taken a stretching  force such that the  elongation predicted by the unconstrained  WLC model is 
$  \bra\, z (L)\,\ket=1.16 L_0$  when $L$  takes  its maximum value. It is then clear that such a molecule,  stretched according to the usual
WLC model, could not fit between the plates. When $L $ approaches its maximum  value $1.5\, L_0  $, the elongation  $ \bra\, z (L)\ket >$   
is no  longer increasing  like  $L$  but tends to its maximum possible value $L_0$; 
at the same time the slope of the hill exhibited by $P(z,L)$ is  becoming  steeper. This
indicates a decrease of the longitudinal  fluctuations of the terminal monomer. }
\label{fig3}
\end{figure}
This section is devoted to the analysis of the results of the transfer-matrix iteration 
 for  molecular chains  having  cristallographic  lengths within  the interval  $ 20 A \leq  L \leq 30 A $ when they are confined between two plates 
  separated by  a distance $L_0 = 20 A$.   The stretching  force is specified  by taking  the value 
  $ \alpha =F\,A/( k_B\, T) =5 $ for  the reduced  force parameter. 
The unconstrained  WLC  model  would lead to a relative elongation  $ \bra z(L) \ket /L=0.775  $, 
to be  compared with the  ratio  $L_0/L= 2/3\simeq 0.667$.  As a consequence, molecules having  a cristallographic  length  $ L > L_{*}=
L_0 /0.775= 1.29 \, L_0= 25.8 \, A $,  could not fit within  the two plates if stretched according to the WLC model 
prediction.
%%%%%%%%%%%%%%%%%%
\subsubsection{ The terminal monomer statistics.}
%%%%%%%%%%%%%%%%%%%%%%%
 Our iteration procedure provides  us  with 
the the terminal-monomer  distributions $ P(z,L)$ for various molecular lengths  $ L\leq 30~A $. 
At each step we store the probability $ P_{n} \((z(n)\)) $ and  by identifying $ L=n \,b $  we build the  discrete set: $  P(z,n \,b)=
P_{n} (z)$.  The  probability  distribution $ P(z,L)$  is then obtained by interpolation. We  have displayed    in  Fig. 4   a 3-D plot of  
$ P(z,L)$. It allows us to follow in a continuous   way  how the DNA molecule manages  to satisfy the two-plate space constraints. 
When $ 18 \leq   L \leq  20 $ the molecule does not feel   yet the Plate-2 barrier  and the hill ridge, projected upon the 
$(z,L)$  plane,  follows a straight line with a slope  $ \Delta z /\Delta N $ given by 
 the standard WLC model. When  $L$ enters the domain  $ 20\leq   L \leq  30 $  the ridge bends under the 
barrier repulsion to become parallel to the plates. At the same time the hill slope becomes steeper under the combined
effects of the stretching force and the confining plates.

%%%%%%%  FIG5%%%%%%
\begin{figure}
\centerline{\epsfxsize=120mm\epsfbox{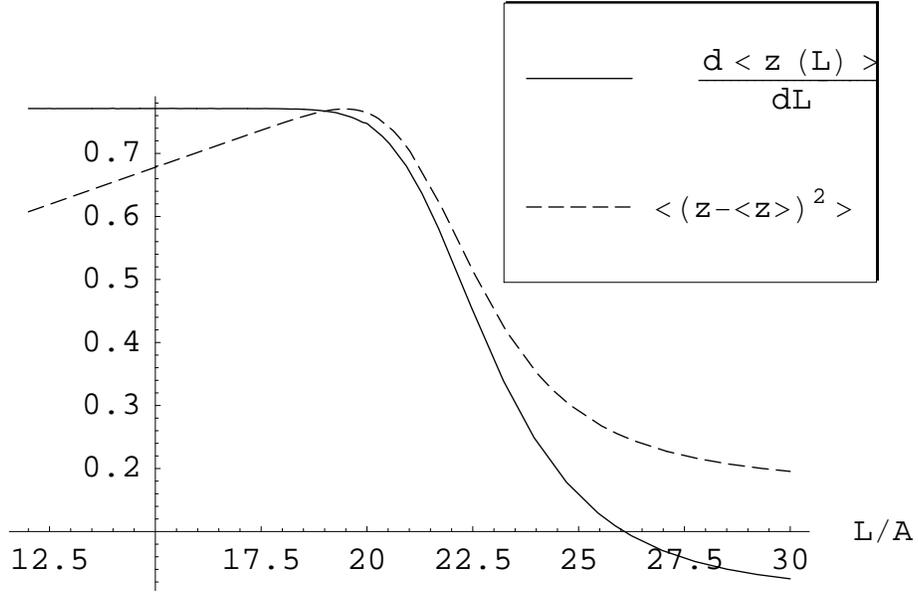}}
%\centerline{\epsfxsize=100mm\epsfbox{FigSC5.eps}}
%\vspace{ -30mm}
\caption{ \small A quantitative analysis of the terminal-monomer  statistics.The  full curve and the dotted curve represent  respectively the
variations
 with  $ L$  of  the two physical quantities,
first, the derivative of the elongation  $ \bra \,z (L)\,\ket $ with respect to   $L$ ,  $\frac{ d \bra\, z (L)\,\ket}{d L}$, 
second,  the longitudinal  fluctuations of the molecular-chain free end:
        $  \Delta \, z^2 = \langle \,( z(L)-\langle z (L)\rangle)^2 \,\rangle $.  One shoud note that 
 both   $ \bra \,z (L)\,\ket $ and  $  \Delta \, z^2(L) $   lose their extensive character when $L\geq L_0$. }
\end{figure}

 In Fig.5  we present  the results  of a  quantitative analysis of the terminal monomer statistics
 described  previously  in a  qualitative way.
   The solid line curve  gives the   derivative of the chain elongation$ \frac{d \bra\, z (L)\,\ket}{ d L }$
as a function of $ L$.
 When $ L $ stays within  the interval $ 12 \,A \leq L \leq  20\, A$,  the rate of variation per unit legnth  $\bra\, z (L)\,\ket $ stays constant 
 as in the unconstrained  WLC model:  the elongation  still behaves as an extensive  quantity. 
We have  verified   that the height of the plateau 
agrees within a few tenths of a percent with the exact WLC prediction quoted above. When the cristallographic length
 of the molecular chain  is going  gradually from $  L_0= 20\, A $ to $L= L_{max}=30  A$, the
 elongation derivative with respect to $L$  starts a rather sharp decrease   towards $0$. 
The elongation  $ \bra \, z(L)  \,\ket$  is no longer extensive and goes slowly  to $ 19 A \simeq L_0.$
The dotted line of Fig.5   gives,  as a  function of  $L$,  the  mean square free-end fluctuations 
  along the  stretching force direction  $  \Delta \, z^2 (L)= \langle ( z(L)-
\langle z(L) \rangle)^2 \rangle $.  
If  $  L \leq L_0 $, then, $ \Delta \, z^2 (L)$   increases linearly  
with  $L$    and the slope  stays very close to that  predicted by the WLC model.
 When  $L$  varies from  $L_0$  to $ L_{max}$,    
 $  \Delta \, z^2 (L) $  undergoes  a sharp decrease and reaches a final  value   
ten  times below the  prediction of the  unconstrained WLC  model. 
 
We would like to  stress  that a  basic physical feature of 
our  model is  the Internal Confinement  ( IC),  which simply means that  all the monomers are
 constrained  practically   to stay 
  within the space  domain:   $ 0 \leq z_n \leq  L_0$. 
We will find  helpful, later on,   to consider also  models involving an External Confinement (EC): 
 the confining potential  $ V(\vr )$ is acting only upon the terminal monomer, for instance via an attached bead, so that the internal  monomers are
no longer spatially constrained.  This type of confinement can be implemented  by adding  to the elastic density    (\ref{elasdens}) the velocity
dependent  contribution $ \dot{\vr}\cdot \nabla_{ \vr} \,V( \vr)$.
 We have calculated  $ \frac{d \bra\, z (L)\,\ket}{ d L }$ and $  \Delta \, z^2L $ 
 versus $L$ using the EC version of the WLC model with the confining potential of Fig.2. These two quantities exhibit a decrease within 
the interval: $ L_0 \leq  L \leq L_{max} $, but  they stay well above the values  given in Fig.5. For instance, when $L$  goes   from
$25 \,A$  to $ L_{max}= 30 A, \, \frac{d \bra\, z (L)\,\ket}{ d L }$ decreases only from $ 0.4 $  to $ 0.3$ and $  \Delta \, z^2 (L)$ from 1 to 0.6.

%%%%%%%  FIG6%%%%%%
\begin{figure}
\centerline{\epsfxsize=120mm\epsfbox{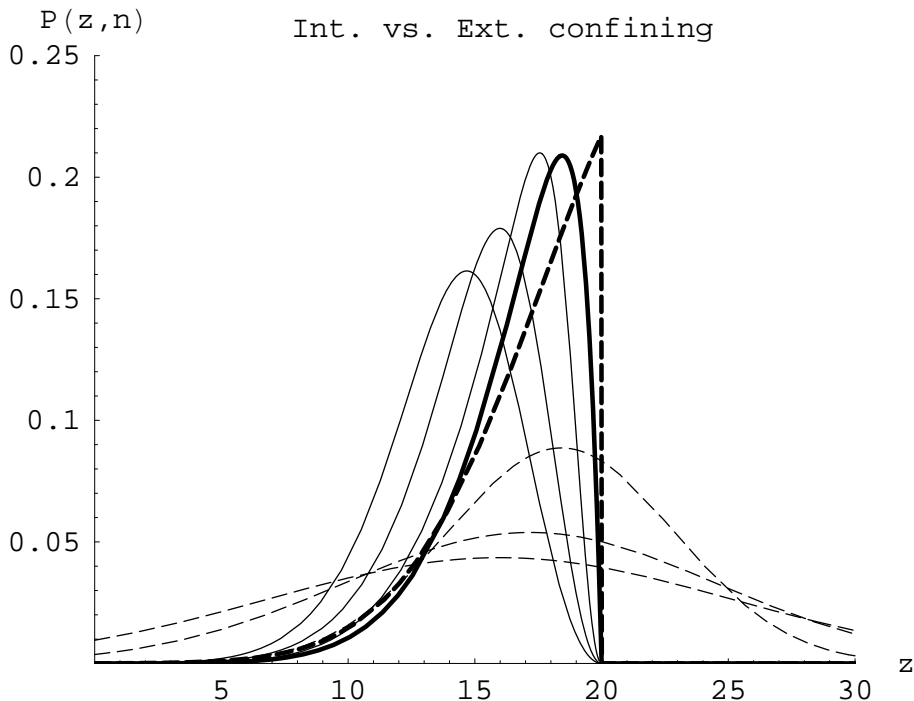}}
%\centerline{\epsfxsize=100mm\epsfbox{FigSC6.eps}}
%\vspace{ -30mm}
\caption{Differences  between   the internal monomer statistics depending upon the type of the Gaussian confining model.
In the  internal-confining (IC) model, all the monomers feel the confining
potential. This contrasts with the   external-confining(EC) model where the confining plate, at $L_0 = 20\, a$, is 
  acting only upon the terminal monomer. 
The solid  and dashed curves stand respectively  for the IC and EC  probabilitiy
 distributions $P(z,n)$ for $L_{int}= n\,b =25 \,a ,27 \, a , 29\,a $, in that order 
from left to right; the thick lines are relative to the terminal monomer.  The
flattening of the terminal segment -  $20 \, \% $ of the monomers- against the  
 repulsive barrier is conspicuous  in the IC  model while nothing similar is visible in the EC case, where internal monomers are allowed to
wander beyond the plate.  }  
\end{figure}

\subsubsection{ The internal monomer statistics. IC versus EC model}
All the curves we have presented  so far concern the statistics  properties of the terminal  monomer
of molecules having their crystallographic length within the range $ 10 A \leq L \leq L_{max}$. 
In  the case  $L= L_{max}$,   it is  clearly of interest to investigate
 the  statistics  of internal chain-segments  with monomer numbers $n$
within  the range    $ N- \Delta N \leq  n \leq  N $, where $ N= L_{max} /b $ and $ \Delta N=L_0/b$. 
The internal monomer statistics  properties  can  be obtained directly within the WLC model,   
but at the price of non trivial modifications of   the Mathematica  codes  used  to get  the results displayed  in Fig.3, 4, 5.
 This would lead  to  developments falling outside the scope  of the present paper. 
 Nevertheless, a rather good  physical understanding of what is going on near
the confining barrier can be obtained within    the  "Gaussian "model, often used to discuss flexible polymers. 
 It  is defined by the   elastic linear density: 
 ${\cal E}^{gaus}=  \frac{ 1}{2\,a}\,( \dot \vr)^2+ V(\vr)$; 
  $ a^{-1} $   is  proportional to the rigidity of the harmonic potential acting between  nearest-neighbour effective monomers.
If we indentify  $  V(\vr)$   with the potential $V(z)$ of  Fig. 1, the internal monomer  probability distribution, $ P_{n}^{int}(x_n,y_n,
z_n)$,  factorizes into three independent distributions relative to  each component. (we remind that $n$ is the effective 
nonomer number related to the coarse-grained $s$ variable  by $s=n \, b$.)   In the continuous limit  $ b/a \ll 1$ ,   our simple statistical
model is easily  solved by exploiting the analogy with  a QM problem (see Appendix); it  involves   the following   simple Hamiltonian:
\be \widehat{H}_z = \widehat{H}_{z \,0 }+ V(z) \;\;  , \;\;\widehat{H}_{z \,0 }= -\frac{a}{2}\,\frac{\partial^2 }{\partial z ^2} \; .   \ee
The internal-monomer  probability distributions relative to  the IC model  read as follows: 
\bea
P_{IC}(n,z_n)&=&\int dz_N \,  \exp(f \,z_N)\bra z_N \vert \exp\(( -b\,(N-n) \widehat{H}_z \))\vert z_n \ket \times \nonumber \\
&  & \bra z_n \vert \exp( -b \,n \widehat{H}_z)\vert z_0 \ket . 
\label{Pzint}
\eea
On the other hand, the formula giving $P_{EC}(n,z_n)$   is obtained  by  perfoming in the r.h.s. of the above equation the following
replacements:
$$  f z_N   \longrightarrow f z_N -b V(z_N) \; \; ,\;\;  \widehat{H}_z   \longrightarrow  \widehat{H}_{z \,0 }.  $$
 The  QM problems  associated  with both    $ \widehat{H}_{z \,0 }$ and $\widehat{H}_z  $ can be easily 
solved analytically by taking  for $ V(z) $    a square well with a depth $ \gg k_B\,T$ . The results displayed in Fig.6 have been obtained 
with  model parameters  leading to basic physical quantities as close as possible to those  appearing
in the previous    WLC  analysis:   
 $ L_{max}=N\,b = 30 \,a$; $ L_{int}= n \, b = 25\, a, \, 27\,a, \, 29 \, a$; 
 $ L_0 = 2/3\, L_{max}=20\,a$;   $ a\,f = 0.775 $. This last condition  leads to :   $\bra \, z(L) \, \ket /L=0.775 $ in absence of confinement. 

In Fig. 6 the solid and dashed curves refer respectively to the IC and EC model. The two thick lines  are 
relative to the terminal-monomer distributions.   The z-probability distributions for internal monomers with $ L_{int} = 25\, a, \, 27 \, a , \,  29
\, a$  
 -appearing  in that order from left to right  on Fig.6 - exhibit very  striking differences
 between the IC and the EC model.  The IC curves suggest that 
the terminal chain-segment, involving $ 20\%$ of the internal monomers, 
is subject to a progressive flattening  against the repulsive barrier. In contrast
 no such effect is observed  with the EC curves:  the  EC internal  monomers 
 wander rather freely across the repulsive  barrier while the terminal monomer is practically stuck to it, as it should . 

\section{ Perspectives and Possible  Extensions. }
We would like to suggest some  possible applications or extensions of the work presented in this paper.
 \begin{itemize}
\item   The tranverse fluctuations of a single dsDNA molecule attached at both extremities to a plane surface
have been observed recently \cite{Desb}. 
The experimental data have been analysed   with succes by a dynamical model derived from  the WLC  model, ignoring 
the barrier effect of the surface. We believe it is of interest to understand this remarkable result within the  spatially 
constrained WLC  model, using
a closely  related configuration, more easy to handle  than the actual one. Only one extremity of dsDNA segment  
 is assumed to be  anchored  to the surface while a pulling  force  parallel to the surface is acting upon the free end. 
The Boltzmann factor associated with the surface barrier effect will be described by  the  rounded-off  step function introduced
in  Section II.   
\item  
The dsDNA  configurations studied in Section II  
were not fully realistic since we have ignored the spatial obstruction coming from the magnetic (or optical ) tweezers. In practice 
the dsDNA  is attached near the bottom of the spherical bead. In recent micro-manipulation  experiments \cite{Terence} the tendancy is to use
relatively  short   dsDNA segments with 2000 base pairs, corresponding to  $ L = 680 $~nm.  
For  a stretching force $ F=0.3 $~pN, the  molecule   explores the 
surface around its  anchoring point over a  distance of  the order  of the root mean square tranverse fluctuation:
 $\sqrt{ \langle x^2 \rangle }=\sqrt{ \langle z\rangle \,(k_B \,T)/F}\simeq 1.55~A =85$~nm,  
which is about ten times smaller than the bead  diameter. The DNA molecule  does not really feel
the curvature of the bead surface; therefore the bead can be approximated by  its tangent plane. Ignoring the  longitudinal
fluctuations, 
 we  substitute to the obstructive bead a plane barrier lying  at a distance
  $ l= \langle z\rangle $ from the anchoring plate. To compute the bead obstruction  correction, we suggest to follow 
  a method   similar  to that used in Section II, but with one difference: instead of computing the elongation in terms of 
the pulling force,  we shall rather get the force as a function of the elongation. Let   $ F_0$  be the force associated with 
 the given elongation $ l= {\langle z\rangle} $,  when  one  ignores the bead obstruction. The next step will be
to  compute the elongation in presence of the plate, subsitute  for the bead, starting from  a pulling 
force $ F<F_0$ and letting it grow until it reaches the value $ F_1$ where the elongation is back
to its  inital    value $l$. The bead  obstruction  correction to  the  force associated 
with a given elongation $l$ will  then  be estimated as  $ \Delta\, F= F_1-F_0 $. 
 An improvement of the  precision of the computations, with respect to that obtained in  Section II,  
will be necessary. This can be achieved by taking a smaller link length $b$,  say $ b=A/30 $ .

 \item  In references  \cite{BouMez98,moroz,BouMez00} the WLC  model has been generalized to a 
Rod Like Chain  (RLC) Model, involving both bending and twisting rigidities.
This makes possible the study    of   supercoiled dsDNA   entropic elasticity
below the denaturation threshlold.
% Using the transfer matrix formulation of the RLC model to be found in ref.\cite{BouMez00}, 
 One can readily modify  the RLC  model in order to  incorporate spatial constraints.  The  recurrence
relation for  the  partition function $Z_n( z_n,\theta_n,\kappa) $   relative to a supercoiled  DNA molecule, with a  given torque  
$ \Gamma=k_B \,T \kappa $  acting upon  its free end, is obtained by 
 performing  in the r.h.s. of the recurrence relation (\ref{recurZ2plates})  the following replacement:\\ 
 $ {\cal{T}}_{WLC}(\theta_{n+1},\theta_n,f) \rightarrow  {\cal{T}}_{RLC}( \theta_{n+1},\theta_n,f,-\kappa^2) $, where 
$ {\cal{T}}_{RLC}$ is given explicitly in ref.\cite{BouMez00}. The anchoring-plate barrier is  expected to  
have   significant  effects upon the so-called ``hat curves", giving, for a fixed force,  the relative elongation 
versus the supercoiling reduced  parameter $\sigma$. Let us take  the  ``low" force case,  $ F\simeq 0.1$~pN,
where the  RLC model  ``hat curve'' dips steeply into the negative $z$ region when  $ \vert \sigma \vert \geq 0.03$.  This effect is  
 attributed to the creation of plectonem structures which  are allowed to wander in the  $ z <0$ half plane, because of the   
 vanishing of their  stretching-potential energy. Therefore, we can expect important modifications once the spatial constraints, which
forbid the $ z <0$ region, are incorporated.

\end{itemize}
%%%%%%%%%%%%%%%%%%%%%%%
%%%%% APPENDIX%%%%%%%%
%%%%%%%%%%%%%%%%%%%%%
\appendix
\section{Transfer Matrix versus Hamiltonian  methods  for  ds-DNA  Subject to Spatial Constraints.  } 
%%%%%%%%%%%%%%%%%%%%%%%%%%%%%%%%%%%%%%%%%%%%%%%%%%%%%%

\subsection{The auxiliary variable method.}
Our starting point, as in section I, is   the following partition function:
\bea
Z = \int {\cal{D}} \ [ \,\vr \ ] \exp{ \left( -\int_0^L {\cal{E}}(s) ds \))} \,,\\
{\cal{E}}(s) = {\cal{E}}_0(\dot{\vr}^2 ) + \frac{1}{2} \,A\,{ \ddot {\vr} }^2  - \vf \, \dot {\vr}  + V(\vr)\, ,
\label{elquart}
\eea 
with $\dot{\vr} = \frac{d\vr }{ d s}$ and $ \ddot{\vr} = \frac{d^2}{ds^2}(\vr)$.
 The variable  $s$ with  $0 \leq s \leq L  $ results  from a coarse graining  of  the molecular  chain.We  have seen that, 
 for a suitable choice of $ {\cal{E}}_0(\dot{\vr}^2 ) $,    the variable  $s$ coincides, to an arbitrary  precision, with the arc-length  $ s$ 
of a rectifiable curve.

To compute the partition function, the first step involves the discretization of the variable  $s$: $ s_n =n\, b$.  The molecular
chain  is then   represented  by $ N $ elementary links  or effective monomers  with $ N=L/b$.
Assuming that   the effective monomer  length $b $ is  much smaller than the persistence  length   $A$, 
 we replace the derivative by finite differences: 
\be
\dot{\vr} =\frac{ \vr_{n} - \vr_{n-1} }{b}, \hspace{15mm} \ddot{\vr} = \frac{\vr_{n} - 2\vr_{n-1} + \vr_{n-2} }{b^2} \,.
\ee 
The partition  function  is then written   as a multiple integral of a product of    $ N-1$  functions of the  monomer-coordinates $\vr_n$:  
\be
 Z =\int \prod _{n=1}^{N-1} d^3\vr_n  \, \{ \prod _{n=2}^{N}  \exp \((- b  {\cal{E}}_{disc}(n) \)) \}\, Z_{in}( \vr_1,  \vr_0) .
\ee
We are going to  use  a standard  field-theory trick  to eliminate  high order   derivatives  whithout having 
to plug by hand  Dirac delta functions into the functional integral. This involves the introduction   of an auxiliary 
dynamical   variable $\vu$. The starting point  is  the integral  identity, written for  fixed $n$:
\be
\exp\{-\frac{ b \,A}{2}(\ddot{\vr}_n)^2\} \equiv  C(A)^{-1} \,
\int_{-\infty}^{\infty}  d^3\,\vu_n\,\exp\{ -  \frac{b \,A}{2} \(( (
\,{\vu}_n/A+i\ddot{\vr}_n)^2 +\ddot{\vr}_n^2  \))\}\,,
\ee
where $ C (A)  = \int_ {-\infty}^{+\infty} d^3\,{\vu}_n \,  \exp\{-\frac{b } { 2\,A }\,( {\vu}_n + i A\, \ddot{\vr}_n)^2 \}
=( \frac{2\pi\,A}{b})^{\frac{3}{2 \,}} .$ 
Expanding the argument of the exponential we get the final identity:
\be
\exp\{-\frac{b\,A}{2}(\ddot{\vr}_n)^2 \}  \equiv   C(A)^{-1} \,  \int d^3\,\vu_n\,\exp\{-b \,\(( \frac{1}{2\,A}{\vu}_n^2+i\,
\vu_n \cdot \ddot \vr_n \)) \}\,.
\label{basident}          
\ee  
 Ignoring  the overall  factor $ C(A)^{-N}$,  we arrive  in this way  to an  expression of ${\cal{E}}(s) $ involving, now,
 the two variables $ \vr $ and
$\vu$:
\be
{\cal{E}}(s)  \rightarrow {\cal{E}}_0(\dot \vr)^2 + i A \, \vu \cdot \ddot\vr  +
 \frac{1}{2 \,A} \vu^2- \vf \cdot
\dot \vr + V(\vr) \,. 
\ee
We perform the following integration by part:
\be
\int_0^L ds \;\vu\cdot \ddot \vr = \ [\vu \cdot \dot \vr\ ]_0^L -  \int_0^L ds \; (\dot{ \vr} \cdot \dot{\vu}) \,.
\ee
Ignoring   the ``surface term", which  can be estimated  to be of the order of $ 1/N$, 
 we obtain  the new 
expression of the elastic-energy density:
\be
{\cal{E}}^{new}( s)  = {\cal{E}}_0(\dot{\vr}^2) - i \,  \dot{\vu} \cdot
\dot{\vr} - \vf \cdot \dot{\vr} + V(\vr) +  \frac{1 }{2\, A}\, \vu^2\,.
\label{elnew}
\ee
 We have now a ``kinetic term'', which depends only upon  {\it  first-order derivatives}, so that  the construction
of the quantum Hamiltonian from the functional integral  is  now  a  relatively straightforward  affair. However, a price has to be paid:
some extra-work will be required  to elucidate the  physical meaning  of the auxiliary dynamical variable $ \vu$. 
We must now write  the discrete version of $ {\cal{E}}^{new}( s) $. Using the same kind of prescription as in Section I, one gets
easily:
 \be
 {\cal{E}}^{new}_{disc}(n)=   {\cal{E}}_0(\frac{(\vr_{n}-\vr_{n-1})^2}{b^2}) - 
 \frac{i}{b^2}(\vr_{n}-\vr_{n-1})\cdot   (\vu_{n}-\vu_{n-1}) + V( {\vr}_n )+\frac{1 }{2\, A}\, \vu_n^2\,.
 \ee
 We have omitted  the constant  force $\vf $ contribution since it can be easily added up at the end.
We note that the ``kinetic term" 
 \be
 K_0( \vr_{n}-\vr_{n-1},\vu_{n}-\vu_{n-1})
%={\cal{E}}^{new}_{disc}(n)
=   {\cal{E}}_0(\frac{(\vr_{n}-\vr_{n-1})^2}{b^2}) - 
 \frac{i}{b^2}(\vr_{n}-\vr_{n-1})\cdot   (\vu_{n}-\vu_{n-1}),
\label{kinew}
\ee 
 is symmetric under  the exchanges: $ \vr_n \leftrightarrow \vr_{n-1} \; , \;
\vu_n \leftrightarrow \vu_{n-1}$,  while the ``potential''  contribution
 $ U(\vr_n, \vu_n ) =V( {\vr}_n )+\frac{1 }{2\, A}\, \vu_n^2\, $  is clearly not. We can make $
{\cal{E}}^{new}_{disc}(n)$  symmetric by simply doing the replacement $  U(\vr_n, \vu_n ) \rightarrow  \((U(\vr_n, \vu_{n} )+ U(\vr_n,
\vu_{n-1} ) \))/2 $. We not only remain   within  the leeway involved in any discretization procedure, but   we also obtain an improvement since
a trapezoidal integration   is better than  a rectangular one.

 The transfer matrix associated with the  symmetrized energy density $ {\cal{E}}^{new}_{disc}(n+1) $ reads as follows :
\bea 
T^{new}( \vr_{n+1}, \vu_{n+1}\vert\vr_n, \vu_n )& =&  \exp -b \,  {\cal{E}}^{new}_{disc}(n+1)  \nonumber\\ & & 
\hspace{-35mm} =\exp-b \,\{  K_0(
\vr_{n+1}-\vr_{n},\vu_{n+1}-\vu_{n})+U(\vr_{n+1}, \vu_{n+1} )/2 + U(\vr_n, \vu_n )/2 \}.
\eea  
 We define   the transfer operator  $ \widehat{ T}^{new}$    associated with the transfer matrix by writing:
\be
\bra\vr_{n+1}, \vu_{n+1}\vert \widehat{ T}^{new}\vert\vr_n, \vu_n \ket=T^{new}( \vr_{n+1}, \vu_{n+1}\vert\vr_n, \vu_n ).
\ee

 {\it In the following,  $ \widehat{ X}$  
will stand for   linear  operators  acting upon functions  $ f(\vr_n,\vu_n) $ of the dynamical variables. If $  f(\vr_n,\vu_n)$ is 
an eigenfunction, then the associated  eigenvalue will be written without hat: $ X$.} 

 Let us  isolate the ``kinetic'' part of   $ 
\widehat{ T}^{new}$ by  writing:
\bea
 \widehat{ T}^{new} &=& \exp\((-\frac{b}{2}\, \widehat{ U }\)) \;  \hat {T}_0^{new} \; \exp\((-\frac{b}{2}\, \widehat{ U } \)), 
\label{Trop1} \\
 \langle \vr_{n+1}, \vu_{n+1}\vert   \hat{ T}_0^{new}  \vert \vr_n,\vu_n\rangle &= &
\exp\((-b\,K_0(\vr_{n+1}-\vr_{n},\vu_{n+1}-\vu_n)\)) ,
\label{Trop2}
\eea
where $ \widehat{ U } $ is the operator associated with the potential  $  U(\vr_n, \vu_n )$  and 
 $ K_0(  \vr_{n+1}-\vr_{n},\vu_{n+1}-\vu_n) $  the ``kinetic''  elastic-density  term given by eq.(\ref{kinew}).
We note that the ``kinetic" transfer  operator  $\widehat {T}_0^{new} $ is invariant upon space translations: 
$ \vr \rightarrow \vr +\va $  and $ \vu \rightarrow  \vu +\vb $.
 This implies that  $\widehat {T}_0^{new} $  is diagonal within  the momentum basis defined 
 by $  \bra \,\vr \vert  \vp_r \, \ket = \exp i \vp_r \cdot \vr$ and  $  \bra \,\vu \vert  \vp_u \, \ket = \exp \, i \vp_u \cdot \vu $. 
Performing the involved  Fourier transforms, we get  readily the  matrix element  of   $\widehat {T}_0^{new} $  in 
the momentum basis:
\bea
\langle \vp_{\vr_ {n+1}}, \vp_{\vu_{n+1}}\vert   \widehat{ T}_0^{new}  \vert \vp_{\vr_ {n}}, \vp_{\vu _{n}}\rangle&=
& (2 \pi )^6\delta^3 (\vp_{\vr_ {n+1}}-\vp_{\vr_ {n}})\, \delta^3 ( \vp_{\vu_{n+1}}-\vp_{\vu _{n}})\,
 \widetilde{T}_0^{new}(\vp_{\vr_ {n}},\vp_{\vu _{n}} ),  \nonumber \\
\widetilde{T}_0^{new}(\vp_{\vr_ {n}},\vp_{\vu _{n}}) &=& (2 \pi )^3 \, \exp\{- b \((  {\cal{E}}_0( \vp_{\vu_{n}}^2)
+i \, \vp_{\vu_{n}}\cdot \vp_{\vr_ {n}}  \)) \} \,.  \label{T0four2} 
\eea
The two $\delta$-functions appearing above   reflect
 the translation invariance of  the transfer  operator $\widehat {T}_0 $. 
We  note that   $\vp_{\vu_{n}} $ has taken the place of the velocity $ \vv_{n}= {\dot{\vr}}_n $ in ${\cal{E}}_0( {\dot{\vr}}_n ^2)$.
This suggests  that  $ \vp_{u} $  can be identified with the velocity $ \vv$,   as we are going to prove later on.
We have now all what we need to write $ \widehat{T_0}  $  as the exponential of an Hamiltonian  operator  $  \widehat{H_0}  $ 
symmetric  in Fourier space, but non-Hermitian.
Returning to the ordinary space, we introduce {\it the conjugate 
momentum operators } relative to the dynamical variables  $  \vr $ and $\vu $:
\be
\hat{\vp}_{\vr}=-i\nabla_{\vr}  \;  , \;  \hat{\vp}_{\vu} =-i\nabla_{\vu} \, . 
\ee
  Let us remind  that  the transformation $ \tilde{f}( \vp_{\vr})\rightarrow  \vp_{\vr}\tilde{f}( \vp_{\vr}) $, written in the 
Fourier space, reads in ordinary space $ f(\vr) \rightarrow  \hat{\vp}_{\vr}\, f( \vr)$ ( similarly for the variable $\vu$);
so we can write   $ \widehat{T}_0^{new}$  as the  differential operator
$\exp (-b\, \widehat{H}_0^{new} )$,  where 
     $  \widehat{H}_0^{new}  $ is   given, up to an irrelevant additive constant, by the following expression:
\be
\widehat{H}_0^{new}={\cal{E}}_0( \hat{\vp}_{\vu}^2 )
+ i\, \hat{\vp}_{\vu} \cdot  \hat{\vp}_{\vr}.
 \ee
 The transfer operator $ \widehat{T}^{new} $, defined in eq.(\ref{Trop1}), reads then as follows: 
\be 
 \widehat{ T}^{new}=\exp\((-\frac{b}{2}\, \widehat{ U }\)) \;  \exp (-b\, \widehat{H}_0^{new} )\; \exp\((-\frac{b}{2}\, \widehat{ U } \)).
\label{Topvf}
\ee
Using the Cambell-HausDorff formula one can derive the following expression for 
$  \widehat{T}^{new} $ \cite{ilio}:
 \bea
 \widehat{T}^{new}=\exp- b\,\(( \widehat{H}_0^{new}+\widehat{U} +
\frac{b^2 }{24}Ê( 2\, \lbrack \, \lbrack\widehat{U} , \widehat{H}_0^{new}\,\rbrack,\widehat{U}\,\rbrack]-
 \lbrack \, \lbrack\widehat{U} ,
\widehat{H}_{0}^{new}\,\rbrack,\widehat{H}_0^{new} \,\rbrack ) + {\cal{O}}(b^3)\)) .
\label{Tilio}
\eea
The total  Hamiltonian  $\widehat{H}_{new}$  is readily obtained  as the lowest  order term with respect to  b: 
\be
\widehat{H}^{new}=\widehat{H}_0^{new}+\widehat{U}= {\cal{E}}_0( \hat{\vp}_{\vu}^2 )+
 i\, \hat{\vp}_{\vu} \cdot  \hat{\vp}_{\vr} +V(\vr) + \frac{1 }{2\, A}\, \vu^2 .
\ee

Another   interest of the above formula   is to allow a comparison of the  symmetric  transfert operator $ \widehat{T}^{new} $
connecting adjacent effective  monomers along the discretized  chain   with  the exact  evolution operator $ \exp-b \widehat{H}^{new} $
associated with a jump $ \Delta  \,s= b$ along the original continuous chain. The  formula (\ref{Tilio}) shows clearly  that the 
 absolute  uncertanity introduced  by using $ \widehat{T}^{new} $ instead of the exact evolution operator is of
the order of $ b^3$. The  use of an unsymmetrized transfer matrix would have led to an  absolute  uncertainty  of the order $b^2$.
%%%%%%%%%%%%%% subSection  2.2     %%%%%%%%%%%%%%%%%%%%%%%%%%
\subsection{  Physical interpretation of  the auxiliary  variable $\vu$ : the final form of
the Hamiltonian. } 
%%%%%%%%%%%%%%%%%%%%%%%%%%%%%%%  
In the above derivation of  $\widehat{H}^{new}$, the auxiliary variable  $\vu$ and its conjugate  momentum  operator $ \hat{\vp}_{\vu}$
 have  been introduced in a rather formal  way. We  would  like to show    that 
the operator $ \hat{\vp}_{\vu}$ has a well defined physical interpretation. The fact that  it 
  replaces  $ \dot{\vr }$  in going  from ${\cal{E}}_{new}$ to $ \widehat{H} $, suggests
  that these two quantities are identical. 

To get an explicit proof,   it is   convenient to 
introduce the ``Euclidian Heisenberg" operator relative to the monomer coordinate  along the chain:
 \be 
\hat{\vr}(s)=\exp(s \,\widehat{H} ) \,\vr\exp(-s \,\widehat{H} ).
\ee
 These operators  enter in the  computation of  
 the correlation function of the two-momomer coordinates which is given by the standard formula \cite{Parisi}:
\bea
&& C\((r_i(s_1),r_j(s_2)\))= \nonumber \\
&&\mathrm{Tr}\(( (\exp-L \, \widehat{H})\,\hat{r}_i(s_1) \, \hat{r}_j(s_2)  \))
/\mathrm{Tr}\(( \exp-L \,\widehat{H}\)). \; \; \;
\label{correl}
\eea
  The  two-monomer correlation function relative  to the velocities is obtained by  taking the partial derivatives
of  $ C\((r_i(s_1),r_j(s_2)\))$ with respect to $s_1$ and  $s_2$.
 We have  then to compute the  s-derivative of   $ \hat{\vr}(s) $:
 \be 
 \hat{\vv}(s)= \frac{d\, \hat{\vr}(s)}{d\,s}= \exp(s \,\widehat{H} )\, [ \widehat{H},\vr ]\, \exp(-s \,\widehat{H} )= \hat{\vp}_{\vu}(s)\,.
\label{puvidentity}
 \ee
Going to the limit $ s \rightarrow 0$   the operator $ \hat{\vp}_{\vu} $
 is just the Euclidian   velocity operator  $\hat{\vv}=  [ \widehat{H},\vr]$.
 It is clear that the above computation  can be applied to more general situations  where   $\hat{r}_j(s_2)$  is replaced  by  any 
physical operator. Let us choose, for instance,  the  Unit operator {\large 1\hspace{-1.5 mm}I}. One obtains  readily  
 a simple  physical confirmation of the  $   \hat{\vp}_{\vu} \Leftrightarrow \hat{\vv}   $ identity
 through the following relations between thermal averages:
 \be
   \langle  \dot{\vr}(s) \rangle =\langle   \vv(s) \rangle =\langle   {\vp}_{\vu}(s) \rangle. 
\ee
In all the explicit computations to be  performed later on ,  we shall use a functional  basis diagonal   with respect to
 the velocity vector  operator   $\hat{\vv} \Phi  = \vv \,\Phi $, 
where the real vector $\vv $ stands for the three corresponding  eigenvalues.
We can use a more familiar language by   working with  the Fourier transforms
 of the wave functions:  $ \Psi (\vu)=\int  d^3\vv\exp( i \vu\cdot \vv)\,\Phi(\vv)$.  The two linear
transformations 
$ \Psi (\vu)\rightarrow \hat{\vp}_{\vu}\,\Psi (\vu)$  and $\Psi (\vu)\rightarrow\vu\,\Psi (\vu)$
read respectively in Fourier space: $ \Phi (\vv)\rightarrow \vv\,\Phi (\vv)$  and $\Phi (\vv)\rightarrow  i\,\nabla_{\vv}\,\Phi (\vv)$.
 We get the  final expression for the  ``physical" Hamiltonian  $\widehat{H}$   by performing upon   $\widehat{H}^{new}$  
 the replacement  $   \hat{\vp}_{\vu}  \rightarrow \hat{\vv}$ and $  \vu  \rightarrow  i\, \nabla_{\vv}  $:
\be
    \widehat{H }= -\frac{1}{2\, A} {\nabla_{\vv}}^2 +{\cal{E}}_0( \vv^2)+\vv \cdot ({\nabla}_{\vr} -\vf) +V(\vr)\,.
\label{hamfinal}
\ee 
 As noted
before,  $\widehat{H }$ is not an  Hermitian  operator, but if 
$ V(\vr)$ is an  even function of $\vr$,  it is self-adjoint with respect to the functional scalar   product:
 $  (\Phi_1 \vert \Phi_2)=\int d^3\vv\, d^3\vr \, \Phi_1 (\vv,-\vr)\,\Phi_2( \vv,\vr)$.
%%%%%%%%%%%%%%%%%%%%%%%%%%%%%%%%%%%%%%%%%%%%%%%%%%%%%%%%%%%%%%%%%%%%%
\subsection{ The   Spatially Constrained WLC Hamiltonian.} 
In order to apply the present formalism  to the WLC  model with spatial constraints,   we have to take an  adequate
functional form   for  ${\cal{E}}_0(\vv^2)$.  As it was shown previously in section I, the solution is rather simple: 
one introduces the small length $ \delta b$ such that 
$ \delta\,b/b \ll 1$ and onchooses  for  $ {\cal{E}}_0(\vv^2) $ the following expression:
  \be
{\cal{E}}_0(\vv^2)= b\frac { ( \vv^2-1)^2 }{2 \,\delta b^2  }\,.
\ee
Introducing  the above choice  of ${\cal{E}}_0(\vv^2)$ in  the final expression  for the  Hamiltonian  $\widehat{H}$ 
given by equation (\ref{hamfinal}),  we get the Hamiltonian $\widehat{H}(\delta b )$ :
\be
   \widehat{H}(\delta b )= -\frac{1}{2\, A} {\nabla_{\vv}}^2 +b\frac { ( \vv^2-1)^2 }{2 \,\delta b^2}+
\vv \cdot ( {\nabla}_{\vr} -\vf) +V(\vr)\,.
\ee
Performing the trace  upon   $ v= \vert \vv \vert $  upon  the  transfer operator $ \exp-b \, \widehat{H}(\vv, \delta b ) $,    
only  the value $ v=1$  
 does  contribute in the limit  $\delta \,b \rightarrow 0$,   so that $\vv $  coincides with the unitary tangent vector $\vt $
appearing in the WLC model. We arrive   in this way  to the  
    WLC  Hamiltonian adequate for spatially constrained dsDNA \cite{Mag,Gom,Burk93,Burk97, Burk01,kier,saito,fried,Yama,helf, mors}:
\be
{\widehat{H}}_{SCWLC}=  -\frac{1}{2\, A} {\nabla}_{\vt}^2 - \vf \cdot \vt +\vt\cdot {\nabla}_{\vr}+V(\vr)=
 {\widehat{H}}_{WLC}(\vf) +\vt\cdot {\nabla}_{\vr}+V(\vr)\,.
\label{hamWLCext}
\ee

If  $ V(\vr) =0$, it is easily seen that  ${\widehat{H}}_{SCWLC}$ is diagonal in the momentum space.
 Perfoming  the  trace with respect to $ \vr $  upon   the exact  transfer matrix 
  is equivalent,   within the momentum  basis,  to take the limit $ {\vp_r} \rightarrow 0$.
 In this way,  the standard WLC model Hamiltonian is immediately  recovered  \cite{Yama}.
%%%%%%%%%%%%%%%      subsection 3.3 %%%%%%%%%%%%%%%%%%
\subsection { The transfer matrix deduced  from  the  SCWLC Hamiltonian.}
%%%%%%%%%%%%%%%%%%%%%%%%%%%
It is of interest   to compare  the symmetric transfer matrix obtained
   from  the Hamitonian ${\widehat{H}}_{SCWLC}$,  by  an equation similar to  (A19),
to the one derived  in Section I, directly  
 from the  the partition functional integral  of eq.(\ref{intfuncZ}).
  
 To proceed, we write   the  Hamiltonian  ${\widehat{H}}_{SCWLC}$  of eq.(\ref{hamWLCext})  
 under the following  compact form:
\be
{\widehat{H}}_{SCWLC}= {\widehat{H}}_{WLC}(\vf-i \,\hat{\vp}_{\vr}  ) +V(\vr)Ê\, . 
\ee
Using  eq.( \ref{Tilio}),  the exact transfer operator  $  \exp (- b {\widehat{H}}_{SCWLC}) $  can be written, 
 up to corrections of the order of $ b^3$, as follows :
\be
 {\widehat{ T} }_{SCWLC}=\exp\((-b\,\frac{\widehat{V}}{ 2}\))\,\exp\((-b\,{\widehat{H}}_{WLC}(\vf-i \,\hat{\vp}_{\vr}  )\))
\,\exp\((-b\,\frac{\widehat{V}}{ 2}\)) \,.
 \ee
We can write the recurrence relation obeyed by  the  partition function $Z_n ( {\vr}_n,{\vt}_n)$  
using the transfer matrix associated with $ {\widehat{ T} }_{SCWLC} $:
\bea 
Z_{n+1} ( {\vr}_{n+1},{\vt}_{n+1})&=&\int d^3 \vr_n \,\int d^2\,\Omega(\vt_n )\,
\exp\((-b\,\frac{V({\vr}_{n+1})+V({\vr}_{n})}{ 2} \))\nonumber\\
& &
 \langle\vr_{n+1},  \vu_{n+1}\,  \vert\, 
 {\widehat{T}}_{WLC}(\vf-i \,\hat{\vp}_{\vr}  )\vert\vr_n,\vu_n\rangle \, Z_n( \vr_n, \vt_n)\,.
\label{recurZ}
 \eea
Let us, first,   evaluate the   matrix element  of ${\widehat{T}}_{WLC}(\vf-i \,\hat{\vp}_{\vr}  )$, which 
 is diagonal in the momentum  basis. The corresponding diagonal element  
is  just  the WLC-model  transfer matrix given  in equation (\ref{trmatWLC})  for   the case of a complex force
 $\vf -i {\vp}_{\vr}$. We get  in this way an  intermediate expression  of the ${\widehat{T}}_{WLC}(\vf-i
\,\hat{\vp}_{\vr}  )$ matrix  element, within the coordinate space, written  as a Fourier integral:
\bea 
\langle\vr_{n+1},  \vt_{n+1}\,  \vert\, 
{\widehat{T}}_{WLC}(\vf-i \,\hat{\vp}_{\vr}  )\vert\vr_n,\vt_n\rangle &=& 
\int \frac{ d^3\,{\vp}_{\vr }}{ (2 \,\pi)^3}\exp i\,{\vp}_{\vr}\cdot \(( {\vr}_n  -{\vr}_{n+1} +b\,\frac{ {\vt}_{n+1}+{\vt}_n}{2} \))\times
\nonumber \\
& &  \langle \vt_{n+1}\,  \vert\, {\widehat{T}}_{WLC}(\vf )\vert\vt_n\rangle .   
\eea
 The integration over ${\vp}_{\vr }$ gives  a Dirac  $\delta$-function   which 
is similar  to that introduced by hand in Section I:
\be
\langle\vr_{n+1},  \vt_{n+1}\,  \vert\, 
{\widehat{T}}_{WLC}(\vf-i \,\hat{\vp}_{\vr}  )\vert\vr_n,\vt_n\rangle= \delta ^3\(({\vr}_{n+1} -{\vr}_{n}  - b\frac{
{\vt}_{n+1}+{\vt}_n}{2} \)) \langle \vt_{n+1}\,  \vert\, {\widehat{T}}_{WLC}(\vf )\vert\vt_n\rangle \,.
\ee 
Performing  the integration  over  $\vr_n $  leads   immediately  to the  partition-function recurrence relation:
\bea 
Z_{n+1} ( {\vr}_{n+1},{\vt}_{n+1})&=&\exp\((-\frac{b}{2}\,V({\vr}_{n+1})\))\int \,d^2\,\Omega(\vt_n)
\,\exp\lbrace-\frac{b}{2}\,V\(({\vr}_{n+1}-\frac{b}{2}( {\vt}_{n+1}+{\vt}_n  )\)) \rbrace \nonumber\\
& & \langle \vt_{n+1}\,  \vert\, {\widehat{T}}_{WLC}(\vf )\vert\vt_n\rangle
\, Z_n\(( \vr_{n+1}-\frac{b}{2}( {\vt}_{n+1}+{\vt}_n  ), {\vt}_n \))\,.
\label{recurZbis}
 \eea
\subsection { Relation berween the symmetric and asymmetric  transfer matrices.}
In principle, one can build  a  numerical iteration procedure  from the above  recurrence relation. 
It will turn out to be more complex and physically less transparent than the one which is actually  used 
in the present paper. We are going to describe  the simplifications to be made on eq.(\ref{recurZbis}) in order to recover
 the recurrence relation (\ref{recurZtrvf}) derived directly in Section I, by starting from the Boltzmann functional
integral.
 \paragraph{Basic assumptions.}
 In order to  justify this procedure, we shall  assume that 
 the rates of variation  with $ \vert \vr \vert $, both of the  potential  $ V(\vr)$   and the 
 starting  partition function $Z_0( \vr,\vt  )$,  are  at most  of the order  of the inverse of the persistence length
$A$, which is  assumed to be at least  ten times  larger than the length  $b$ of the elementary link.  
More precisely we shall impose the two constraints:  $Z_0( \vr,\vt  )^{-1} \, \nabla_{\vr} \,Z_0( \vr,\vt  )\sim A^{-1}$ and 
$V( \vr)^{-1} \, \nabla_{\vr}\, V(\vr  )\sim A^{-1}$.
In the computations presented in Section II, these constraints were indeed satisfied.   
One can prove by recurrence that these estimates   hold  also for    $Z_n(  \vr_n,\vt_n  )$.This last point
can  be  verified   by looking at  the  the probabilitity  distributions 
$P(z,L)$ plotted on Fig. 5, where $A$ is taken as the length unit.
 This  will  allow us to make two
important simplifications:
\paragraph{Simplification 1.}
 The two  Boltzmann factors involving the potentiel  $ V(\vr) $ can be replaced by a single one, namely:
$ \exp-b \,  V({\vr}_{n+1} ).$ For the  confining potential $V(z)$  used in this paper,  we have 
 found that the corresponding  Boltzmann factor corrections are  
  about  $ 2 \%$; they are  localized near the  two plate barriers in   narrow regions of width $ \sim A$. Their
 overall effect is expected  to  be  somewhat smaller,  since the width  of the potential well is  $ > 20 A$.
  \paragraph{Simplification 2.}
  The second simplification is  a  change of the monomer coordinate in  
the partition function $ Z_n  $  appearing in  the r.h.s of eq.(\ref{recurZbis}): 

$ \vr^{b}_{n+1}={\vr}_{n+1} -\frac{b}{2}( {\vt}_{n+1}+{\vt}_n) \rightarrow
\vr_{n+1}^a=   {\vr}_{n+1}-b \, {\vt}_{n+1}. $
It is to convenient rewrite $\vr_{n+1}^a$ and   $\vr^{b}_{n+1}$ in terms of $ \Delta\vt_n=( \vt_{n+1}- \vt_n) $:
\be
\vr_{n+1}^a= {\vr}_{n+1}- b\, \vt_n -b\, \Delta\vt_n \hspace {6mm} \rm  {and } \hspace {6mm }\vr_{n+1}^b= {\vr}_{n+1}- b\, \vt_n -b/2\,
\Delta\vt_n. 
\label{ra&rb}
\ee
We get expressions for the  probability  distributions  $ P^a( \vr_{n+1} )$ (in the limit $ \vf \rightarrow 0 $) 
by integration over $\vt_{n+1}$ of the two sides of eq.(\ref{recurZtrvf})   and  then  exchange of the   integration order:
\bea 
  P^a( \vr_{n+1} )&=&\exp\((-b\,V({\vr}_{n+1})\))\, \int d^2\,\Omega(\vt_n )\, \int d^2\,\Omega(\vt_{n+1}) 
    \times \nonumber \\
& &  \exp\, \(( - \frac{b}{2 \, A}  \Delta\vt_n^2 \))Z_n ( \vr_{n+1}-b \, \vt_n -b \Delta\vt_n ,\,\vt_n ). \nonumber 
\eea
We now make a  first  order Taylor expansion of $ Z_n$ with respect to $ b \Delta\vt_n$ and perform the
 Gaussian average over $\Delta\vt_n $  using  $\vt_{n+1}$ as integration variable. We arrive in this way to 
a formula useful for our purpose:
\be
P^a( \vr_{n+1} )=\exp\((-b\,V({\vr}_{n+1})\))\, \int d^2\,\Omega(\vt_n )    
\(( 1- \frac{b^2}{A}\, \nabla_{\vr_n}\cdot \vt_n \))Z_n ( \vr_{n+1}-b \, \vt_n ,\,\vt_n ).
\ee
The formula for $ P^b( \vr_{n+1} )$ is obtained
 by mutipliying by $ 1/2 $ the $ b^2/A$ term, as it follows immediately from eq.(\ref{ra&rb}.)
Using the above basic assumptions, one finds that  the difference $P^a( \vr_{n+1})- P^b( \vr_{n+1} )$  is  $ \sim \frac{1}{2}\,( b/A)^2$
and  this reflects the lack of symmetry of the transfer matrix used in Section I.
 By  implementing   the simplifications  1. and 2. in equation (\ref{recurZbis}),   we  recover immediately
the partition function recurrence relation derived in Section I (eq.(\ref{recurZtrvf}):
\be 
Z_{n+1} ( {\vr}_{n+1},{\vt}_{n+1})=\exp\((-b\,V({\vr}_{n+1})\))   \int d^3\vt_n
 \langle \vt_{n+1}\,  \vert\, {\widehat{T}}_{WLC}(\vf )\vert\vt_n\rangle  
Z_n( \vr_{n+1}-b\, {\vt}_{n+1},\vt_n )\,.
\label{recurZfinal}
 \ee
\section*{Acknowledgments}
It is a pleasure to thank D. Bensimon, M.-A. Bouchiat and J. Iliopoulos for their careful reading of this manuscript and valuable suggestions. 

Laboratoire de Physique Th\'eorique de l'Ecole Normale Sup\'erieure is a Unit\'e Mixte du Centre National de la Recherche Scientifique
et de l'Ecole Normale Sup\'erieure (UMR 8549).
 

\begin{thebibliography}{30}
\bibitem{busta}
  C . Bustamante, Z. Bryant and S. Smith, Nature {\bf 421}, 423-427 (2003).
\bibitem{allem} 
J.F. Allemand, D. Bensimon and V. Croquette,  Curr. Op. Structural Biology {\bf 13}, 266-274 (2003). 
\bibitem{smith}
S.B. Smith, L. Finzi and C. Bustamante, Science {\bf 258}, 1122 (1992).
\bibitem{perk}
 T.T. Perkins, S.R. Quake, D.E. Smith and S. Chu, Science {\bf 264}, 8222 (1994).
\bibitem{strick}
T.R. Strick, J.-F. Allemand, D. Bensimon, A. Bensimon and V. Croquette,
Science {\bf 271}, 1835 (1996).
\bibitem{fixman}
M. Fixman and J. Kovac, J. Chem. Phys. {\bf 58}, 1564 (1973). 
\bibitem{marsig}
J.F. Marko and E.D. Siggia,  Science {\bf 265}, 506 (1994).
\bibitem{bouchbiophys} 
 C. Bouchiat, M.D. Wang, S.M. Block, J.-F. Allemand and V. Croquette, Biophys. J. {\bf 76}, 409 (1999). 
\bibitem{moroz}
  J.D. Moroz and P. Nelson, Proc. Natl. Acad. Sci. USA  
{\bf 94}, 11418 (1997), Macromolecules {\bf 3}, 6333 (1998).
\bibitem{BouMez98}
C. Bouchiat and M. M\' ezard, Phys. Rev. Lett. {\bf  80}, 1556 (1998).
\bibitem{BouMez00}
C. Bouchiat and M. M\' ezard,  Eur. Phys.  J.E. {\bf 2}, 377-402 (2000).
\bibitem{Mag}
A.C. Maggs, D.A. Huse and S. Leibler  Europhys. Lett.  {\bf 8}, 615 (1988).
\bibitem{Gom}
G. Gompper and T.W. Burkhardt, Phys. Rev. A  , 6124 (1989).
\bibitem{Burk93} 
T. W. Burkhardt, J. Phys A {\bf 26}, L1157-L1162 (1993).
\bibitem{Burk97} 
T. W. Burkhardt, J. Phys A {\bf  30}, L167-L172 (1997).
\bibitem{Burk01}D.J. Bicout and J.T. W. Burkhardt, J.Phys. A{\bf 34}, 5745 (2001).
\bibitem{kier}  
J. Kierfeld and R. Lipovwsky, J. Phys. A {\bf 38} L155-L161( 2005).
\bibitem{saito}  
N. Saito, K. Takahashi and Y.Yunoki, J.  Phys. Soc. Japan, {\bf 22},219-236 (1967)  
\bibitem{fried}
Karl L. Fried, J. Chem. Phys.  {\bf 44}, 1153-1463 (1971).
\bibitem{Yama} 
H. Yamakawa, in {\it Helical Wormlike Chains in Polymer Solutions} (Springer-Verlag, New-York, 1997), chapter {\bf 3}, 19-47.
\bibitem{helf}
E. Helfand and Y. Tagami, J. Chem. Phys. {\bf 56}, 3592-3601,(1971)
\bibitem{mors}
D.C. Morse and G.H. Fredrickson, Phys. Rev. Lett. {\bf  73},  3235 (1994).
\bibitem{Parisi}
G. Parisi, in {\it Statistical Field Theory in Advanced Book Classics}, (Perseus Books 1998).
\bibitem{TM} 
  The transfer matrix  method   was applied  to the solution of the one- and two-dimensional Ising Models,
 by K. Huang in {\it Statistical  Mechanics}, (John Wiley and Sons, New York, 1988). 
A good pedagogical introduction  to the formalism  used in the present paper is found in reference \cite{Parisi}. 
The authors of  reference \cite{kier}, working  within a continuous formalism, call transfer matrices
  mathematical objects  which are usually 
known under the name of Green functions: see equations (3) and (7). 
\bibitem{ilio}
 J. Iliopoulos,  Private communication.
\bibitem{Desb} 
A. Crut, D. Lasne, J.-F.  Allemand, M. Dahan and P. Desbiolles, Phys. Rev.  E {\bf  67}, 051910  (2003)
\bibitem{Terence}
A. Revyakin, R. H. Ebright, and  T. R.  Strick,  Nature Methods, {\bf 2}, 127-138 (2005)
\end{thebibliography}
\end{document}